\documentclass[runningheads,fleqn]{svmult}

\usepackage{makeidx}%,showidx}   % allows index generation
\usepackage{graphicx}  % standard LaTeX graphics tool
\usepackage{subeqnar}  % subnumbers individual equations
\usepackage{multicol}  % used for the two-column index
\usepackage{physmult}  % flushleft layout of captions etc.
\usepackage{latexsym}
\usepackage{amssymb}
\usepackage{amsbsy}
\usepackage[mathscr]{eucal}
\newcommand{\ind}[1]{#1\index{#1}}

\def\##1{\langle #1 \rangle}
\let\a=\alpha \let\b=\beta \let\g=\gamma \let\d=\delta
\let\e=\varepsilon \let\z=\zeta  
  \let\l=\lambda \let\m=\mu \let\n=\nu
 \let\s=\sigma
 \let\o=\omega 
\let\ph=\varphi   
\let\O=\Omega

\def\bgamma{\boldsymbol{\gamma}}
\def\mn{{\m\n}}

\def\CC{\mathcal C}
\def\CD{\mathcal D}
\def\CG{\mathscr G}
\def\CH{\mathcal H}
\def\CL{\mathcal L}
\def\CO{\mathcal O}
\def\CP{\mathcal P}
\def\CV{\mathcal V}

\def\({\left(} \def\){\right)} 

\def\be{\begin{equation}}
\def\ee{\end{equation}}
\def\bea{\begin{eqnarray}}
\def\eea{\end{eqnarray}}
\def\bp{\begin{picture}(0,0)}
\def\ep{\end{picture}}

\def\nn{\nonumber}

\begin{document}
\title*{Thermal Gauge Field Theories}
\toctitle{Thermal Gauge Field Theories}
% allows explicit linebreak for the table of content
%
%
%\titlerunning{Focusing of a Parallel Beam}
% allows abbreviation of title, if the full title is too long
% to fit in the running head
%
\author{Anton Rebhan}
\institute{Institut 
f\"ur Theoretische Physik, Technische Universit\"at Wien\\
Wiedner Hauptstr. 8-10, A-1040 Vienna, Austria}
\authorrunning{Anton Rebhan}
% if there are more than two authors,
% please abbreviate author list for running head
%
%

\maketitle              % typesets the title of the contribution

\bp\put(263,150){\large\tt TUW 01-16} \ep

{%\renewcommand{\baselinestretch}{0.5}
\footnotesize
{\bf Abstract:} The real- and imaginary-time-formalisms of thermal field
theory and their extension to gauge theories is reviewed. Questions
of gauge (in-)dependence are discussed in detail, in particular
the possible gauge dependences of the singularities of dressed
propagators from which the quasiparticle spectrum is obtained.
The existing results on next-to-leading order corrections to
non-Abelian screening and dispersion laws of hard-thermal-loop
quasiparticles are surveyed. Finally, the role of the asymptotic
thermal masses in self-consistent approximations to thermodynamic
potentials is described and it is shown how the problem of
the apparently poor convergence of thermal perturbation theory
might be overcome.}

\section{Overview}

The theoretical framework for describing ultrarelativistically
hot and dense matter is quantum field theory at finite temperature 
and density. At sufficiently high temperatures and densities,
asymptotic freedom should make it possible to describe
even the fundamental theory of strong interactions,
quantum chromodynamics (QCD), through analytical and mostly
perturbative means. These lectures try to cover both
principal issues related to gauge freedom as well as specific
problems %(and some solutions) 
of thermal perturbation theory
in non-Abelian gauge theories.

After a brief review of the imaginary- and real-time formalisms
of thermal field theory,
%of quantum field theory at finite temperature and density, 
the latter
is extended to gauge theories. Aspects of different treatments 
of Faddeev-Popov ghosts and different gauge choices 
are discussed for general non-Abelian gauge theories,
both in the context of path integrals and in covariant operator
quantization. The dependence
of the formalism on the gauge-fixing parameters
introduced in perturbation theory
is investigated in detail.
Only the partition function and expectation values of 
gauge-invariant observables are entirely gauge independent.
Beyond those it is shown that
the location of singularities 
of gauge and matter propagators, which define screening
behaviour and dispersion laws of the corresponding quasi-particle 
excitations, are gauge independent when calculated systematically. 

At soft momentum
scales it turns out to be necessary to reorganize
perturbation theory such that (at least) the contribution of
the so-called
hard thermal (dense) loops (HTL/HDL) is resummed. The latter
form a gauge-invariant effective action, and their
gauge-fixing independence is verified.
The existing
results of such resummations 
on the modification of the spectrum of HTL quasi-particles
at next-to-leading order (NLO) are reviewed, and a few
cases discussed in more detail, with special attention given
to gauge dependence questions. 

It is shown how screening and damping phenomena become logarithmically
sensitive to the strictly nonperturbative physics of the
chromomagnetostatic sector, with the exception of the case of
zero 3-momentum. In particular the definition of a non-Abelian Debye
mass is discussed at length, also with respect to recent lattice results.

Real parts of the dispersion laws of quasiparticle excitations, on
the other hand, are infrared-safe at NLO. 
However, as the additional collective modes of longitudinal
plasmons and fermionic ``plasminos'' approach the light-cone,
collinear singularities arise, signalling a qualitative change 
of the dispersion laws and requiring additional resummations.
At high momenta only the regular
modes of the fermions and the spatially transverse ones of the
gauge bosons remain and they
tend to asymptotic mass hyperboloids. The
NLO corrections to the asymptotic thermal masses
play an interesting role in a self-consistent
reformulation of thermodynamics in terms of weakly interacting
quasi-particles.
In this application, 
the problem of poor convergence of resummed
thermal perturbation theory resurfaces, but it is shown
that it may be overcome
through approximately self-consistent gap equations.

\index{hard thermal/dense loops|see{HTL}}

\section{Basic Formulae}

Before coming to quantum field theories and gauge theories in
particular, let us begin by recalling some relevant formulae
from quantum statistical mechanics. 

We shall always consider the
\ind{grand canonical ensemble}, in which a system in equilibrium
can exchange energy as well as particles
with a reservoir such that 
only mean values of energy and
other conserved quantities (baryon/lepton number, charge, etc.)
are prescribed through Lagrange multipliers
$\beta=1/T$ and  $\alpha_i=-\b\mu_i$, respectively, where
$T$ is temperature and $\mu_i$ are the various chemical potentials
associated with a set of commuting conserved observables 
$ \hat N_i=\hat N_i^\dagger$ satisfying
$ [\hat N_i,\hat N_j]=0$ and  $ [\hat H,\hat N_i]=0$, where
$\hat H$ is the Hamiltonian.

The statistical density matrix is given by
\be\label{hatrho}
\hat\varrho = { Z^{-1}} \exp\left[-\alpha_i \hat N_i-\beta\hat H\right]\equiv
{ Z^{-1}} 
\exp\Bigl[ -\beta(\underbrace{\hat H-\mu_i \hat N_i}_{ \mbox{$=: \bar H$}}) 
\Bigr]
\ee
where $Z$ is the \ind{partition function}
\be Z = Z(V,T,\mu_i)= {\rm Tr}\; \E^{-\b\bar H}.
\ee
and
the thermal expectation value (ensemble average) of an
operator $\hat A$ is given by
\be
\langle \hat A \rangle = {\rm Tr}[\hat\varrho\hat A] \,.
\ee

When total
energy and particle numbers are extensive quantities\footnote{A
notable exception occurs when general relativity has to be
included.}, i.e.
proportional to the volume $V$, one also has
$\ln Z \propto V$, and since we shall be interested in the
limit $V\to\infty$, it is preferable to define intensive
quantities. The most important one is the
thermodynamic pressure\index{pressure!thermodynamic} 
\be
P = \frac1{\beta V}\ln Z
\ee
which up to a sign is identical to the free energy density $F/V=
-P$ ($F$ is also referred to as {\em the} thermodynamic potential $\Omega$).

Other thermodynamic (or should one say thermo-static?) quantities
can be derived from $P$, such as
particle/charge densities
\be
 n_i=\frac1V\langle\hat N_i\rangle=
{\partial P\over\partial \mu_i},
\ee
energy density 
\be
 \varepsilon=E/V=\frac1V\langle \hat H\rangle=
-\frac1V{\partial\ln Z\over\partial\beta}=
-{\partial(\beta P)\over\partial\beta}
\ee
(at fixed $\alpha_i$), and
entropy density (which will play a prominent role
at the very end of these lectures) \index{entropy}
\bea 
s&=&S/V={1\over V}\langle -\ln\hat\varrho \rangle =
{1\over V}\ln Z+{\beta\over V}\langle \hat H-\mu_i\hat N_i\rangle \nn\\
&=&{\partial P\over\partial T}=\beta (P+\varepsilon-\mu_i n_i)
\eea
In the latter equations one recognizes the familiar
Gibbs-Duhem relation 
\be
E=-PV+TS+\mu_i N_i,
\ee
which explains why $P$ was defined as the (thermodynamic) pressure.
A priori, the hydrodynamic pressure,\index{pressure!hydrodynamic}
which is defined through the spatial components of the
energy-momentum tensor through ${1\over 3}\#{T^{ii}}$,
is a separate object. In equilibrium, it can be identified with the
thermodynamic one through scaling arguments \cite{Landsman:1987uw},
which however do not allow for the possibility of scale (or ``trace'')
anomalies that occur in all quantum field theories with
non-zero $\beta$-function (such as QCD). In \cite{Drummond:1999si}
it has been shown recently that the very presence of the 
\ind{trace anomaly} 
can be used to prove the equivalence of the two in equilibrium.

%\subsection{Relativistic (covariant) generalization}

All the above relations continue to hold in (special) relativistic
situations, namely within the particular inertial frame in which
the heat bath is at rest. In other inertial frames
one has the additional quantity of heat-bath 4-velocity $u^\mu$,
and one can generalize the above formulae by replacing $V=
\int\! d^3x \to \int\limits_{\Sigma\perp u}\! d\Sigma_\mu u^\mu$,
and the operators in (\ref{hatrho}) by
\be
H \to \int\! d\Sigma_\mu T^\mn u_\nu,
\qquad  N_i \to  \int\! d\Sigma_\mu j^\mu_i .
\ee
$\alpha_i$ and $\beta$ are Lorentz scalars (i.e., temperature is by definition
measured in the rest frame of the heat bath). The partition function
can then be written in covariant fashion as \cite{Israel:1981}
\be
Z = {\rm Tr} \left[ \exp \int\! d\Sigma_\mu
\(-\hat T^\mn \beta_\nu - \hat j^\mu_i \alpha_i \) \right],
\ee
where we have introduced an inverse-temperature 4-vector
$\beta^\mu\equiv \beta u^\mu$.

However, in what follows we shall most of the time remain in
the rest frame of the heat bath where $u^\mu=\delta^\mu_0$.

%%%%%%%%% ctp %%%%%%%%%%%%%%%%%%%%
\section{Complex Time Paths}

\index{time path}

With a complete set of states given by the eigenstates
of an operator $\hat\ph$, $\hat\ph | \ph \rangle = \ph | \ph \rangle$,
we formally write
\be
Z={\rm Tr} \left[ \E^{-\beta \hat H} \right] = \sum_\ph \langle \ph | 
\E^{-\beta  \hat H} | \ph \rangle.
\ee
In field theory, we shall boldly use the field operator
$\hat\ph=\hat\ph(t,\vec x)$ in the Heisenberg picture and
write $| \ph \rangle$ for its eigenstates
at a particular time.

Using that the transition
amplitude $\langle \ph_1 | \E^{-\I \hat H(t_1-t_0)} | \ph_0\rangle$ 
[%in Heisenberg picture terminology 
the overlap of states that have eigenvalue $\ph_0(\vec x)$ at time
$t_0$ with states that have eigenvalue $\ph_1(\vec x)$ at time $t_1$]
has the path integral representation
\bea
&&\langle \ph_1 | \E^{-\I  \hat H(t_1-t_0)} | \ph_0 \rangle \nn\\
&&\quad = {\cal N} \!\!\!\!\!\!\!\!
\int\limits_{\ph(t_0,\vec x)=\ph_0 \atop \ph(t_1,\vec x)=\ph_1}
\!\!\!\!\!\!\!\! {\cal D}\ph \; \exp
\I \int_{t_0}^{t_1}\!\! dt \int\! d^3x\, {\cal L}(\ph,\partial \ph)
\eea
we can give a path integral representation for $Z$
that takes care of the density operator
by setting $t_1 = t_0-\I\beta$, and of the trace by integrating over all
configurations with $\ph_1=\ph_0$.

When there is a chemical potential $\mu\not=0$, we have 
$ \hat H\to \bar H= \hat H-\mu  \hat N$ and this implies $L\to L+\mu N$
if there are no time derivatives in $N$, and we have
\be\label{Zctp}
Z = {\cal N} \int {\cal D}\ph \;
\exp \I { \int_{t_0}^{t_0-\I\beta}\!\!\!\! dt} \int\! d^3x\, \bar {\cal L}
\ee
where the path integral is over all configurations periodic in
imaginary time, $\ph({ t_0},\vec x)=\ph({ t_0-\I\beta},\vec x)$.

In this formula, real time has apparently been fixed to $t_0$
and replaced by an imaginary time flow which is periodic with period $\beta$,
the inverse temperature. In equilibrium thermodynamics, this
seems only fitting as nothing depends on time in strict equilibrium.

However, we have not really required time to have a fixed real part. We
have made the end point complex with the same real part, but 
the integration over $t$ in (\ref{Zctp}) need not be a straight
line with fixed $t_0$.
Instead we shall consider a general complex \ind{time path},
and this allows us to define Green functions by the path integral
formula
\be\label{greenfcts}
 \langle \mathrm T_c \hat \ph_1 \cdots \hat \ph_n \rangle
= {\cal N} \int {\cal D}\ph \, \ph_1 \cdots \ph_n 
\exp{\I\int\limits_{\cal C}\! dt\! \int\! d^3x \, \bar {\cal L}}
\ee
where $\mathrm T_c$ now means \ind{contour ordering} along 
the complex time path $ {\cal C}$ from $ t_0$ to $ t_0-\I\beta$
such that $t_i \in {\cal C}$, and $t_1 \succeq t_2 \succeq \cdots \succeq t_n$
with respect to a monotonically increasing contour parameter.

Through quantities like (\ref{greenfcts}) we are not restricted
to time-independent, thermo-static questions, but may also consider
small perturbations of the equilibrium (response theory).

The \ind{time path} introduced in (\ref{greenfcts}) is in fact not
completely arbitrary. Considering spectral representations in
the energy representation leads to the conclusion \cite{Landsman:1987uw}
that $\cal C$ has to be such that the imaginary part of $t$ is
monotonically decreasing. This is a necessary condition for
analyticity; in the limiting case of a constant imaginary part
along (parts of) the contour, 
distributional quantities (generalized functions) arise.

Except for the periodic boundary conditions with regard to the
end points of $\cal C$ (which become anti-periodic for
fermionic field operators and Grassmann-valued ``classical'' fields),
the path integral formula (\ref{greenfcts}) is formally
identical to the familiar one from $T=0$ and $\mu=0$.

Indeed, \ind{perturbation theory} is set up in the usual fashion.
Using the interaction-picture \index{interaction picture}
representation one can derive
\be
\langle  \mathrm T_c \mathcal O_1 \cdots \mathcal O_n \rangle  = {Z_0\over Z}
\langle  \mathrm T_c \mathcal O_1 \cdots \mathcal O_n\, \E^{
\I\!\int_{\mathcal C} \mathcal L_I} \rangle _0,
\ee
where $\mathcal L_I$ is the interaction part of $\cal L$,
and the correlators on the right-hand-side can be evaluated
by a  Wick(-Bloch-DeDominicis) theorem:
\index{Wick theorem} 
\be
%\langle 0|\cdots|0\rangle  \to \langle \mathrm T_c \cdots\rangle $
\langle \mathrm T_c \E^{\I\!\int_{\mathcal C}d^4x\, j\ph}\rangle _0=
\exp\{-{1\over2}\int_{\mathcal C}\int_{\mathcal C}d^4x\,d^4x'
j(x)D^c(x-x')j(x')\},
\ee
where $D^c$ is the 2-point function and this is the only
building block of Feynman graphs with an explicit $T$ and $\mu$
dependence.
It satisfies the \ind{KMS} (Kubo-Martin-Schwinger)
condition
\be\label{DKMS}
D^c(t-\I{\beta})=\pm \E^{-{\mu}{\beta}}D^c(t),
\ee
stating that $\E^{\I\mu t}D(t)$ is periodic (antiperiodic) for
bosons (fermions).

\subsection{Imaginary-Time (Matsubara) Formalism}

\index{imaginary-time formalism}
\index{Matsubara contour}
The simplest possibility for choosing the complex \ind{time path} is
the straight line from $t_0$ to $t_0-\I\beta$, which is named
after Matsubara \cite{Matsubara:1955ws} who first formulated
perturbation theory based on this contour. It is also referred
to as \ind{imaginary-time formalism} (ITF), because for $t_0=0$
one is exclusively dealing with imaginary times. 

Because of the (quasi-)periodicity (\ref{DKMS}), the propagator
is given by a Fourier series
\be
D^c(t) = {1\over -\I{\beta}}\sum_\nu \tilde D(z_\nu) \E^{-\I z_\nu t},
\quad \tilde D(z_\nu) = \int_0^{-\I {\beta}}\!\!\!dt\, D^c(t) \E^{\I z_\nu t}
\ee
with discrete complex (Matsubara) frequencies \index{Matsubara frequencies}
\be
z_\nu=2\pi \I \nu/{\beta}+{\mu}, \quad
\nu\in \left\{ \begin{array}{ll}\mathbb Z & \mbox {bos.} \\
                              \mathbb Z -{1\over2} &  \mbox {ferm.} \\
               \end{array} \right. 
\ee

The transition to Fourier space turns the integrands of Feynman
diagrams from convolutions to products as usually, with the
difference that there is no longer an integral but a discrete
sum over the frequencies, and compared to standard momentum-space
Feynman rules one has
\be
\int {d^4k\over \I(2\pi)^4} \to 
\b^{-1}\sum_\nu \int {d^3k\over i(2\pi)^3},\qquad
\I(2\pi)^4 \d^4(k) \to\b (2\pi)^3 \d_{\nu,0} \d^3(k).
\ee

However, all Green functions that one can calculate in this
formalism are initially defined only for times on $\cal C$, so
all time arguments have the same real part. The analytic continuation
to different times on the real axis is, however, frequently a highly
non-trivial task \cite{Landsman:1987uw}, so that it can be
advantageous to use a formalism that supports real time arguments from
the start.

\subsection{Real-Time (Schwinger-Keldysh) Formalism}

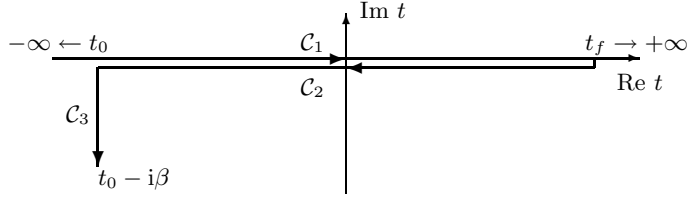
\begin{figure}
\unitlength6mm
\begin{picture}(12,4)(-1,0)
\put(0,3){\vector(1,0){13}}
\put(6.5,0){\vector(0,1){4}}
\put(12.5,2.3){Re $t$}
\put(6.8,3.9){Im $t$}
\put(-1,3.2){\small $-\infty \gets t_0$}
\put(11.8,3.2){\small $t_f \to + \infty$}
\put(1,0.2){\small $t_0-\I\beta$}
\put(5.5,3.2){${\cal C}_1$}
\put(5.5,2.2){${\cal C}_2$}
\put(0.3,1.6){${\cal C}_3$}
\thicklines
\put(1,3){\vector(1,0){5.5}}
\put(1,3){\line(1,0){11}}
\put(12,3){\line(0,-1){0.2}}
\put(1,2.8){\line(1,0){11}}
\put(12,2.8){\vector(-1,0){5.5}}
\put(1,2.8){\vector(0,-1){2.2}}

\end{picture}
\caption{Complex \ind{time path} in the Schwinger-Keldysh real-time
formalism \label{FigRTF}}
\end{figure}

In the so-called real-time formalisms, the complex \ind{time path} $\cal C$ is
chosen such as to include the real-time axis from an initial time $t_0$
to a final time $t_f$. Since we have to end up at $t_0-\I\beta$, this
requires a further part of the contour to run backward in
real time \cite{Schwinger:1961qe,Bakshi:1963}
and to finally pick up the imaginary time
$-\I\beta$. There are a couple of
paths $\cal C$ that have been proposed in the literature. The
oldest one due to Keldysh \cite{Keldysh:1964ud} is shown in Fig.~\ref{FigRTF},
and this is also (again) the most popular one.\index{Keldysh contour}

Clearly, if none of the field operators in (\ref{greenfcts}) has
time argument on $\mathcal C_1$ or $\mathcal C_2$, the contributions
from these parts of the contour simply cancel and one is back to
the ITF.

On the other hand, 
if $t_0 \to -\infty$, and all operators have finite real time arguments,
the contribution from contour $\mathcal C_3$ decouples 
%under normal circumstances 
because from the spectral representation one has for
instance for the propagator connecting contour $\mathcal C_1$ and
$\mathcal C_3$
\be
D^{13}(k,t-(t_0-\I \lambda))= 
\int\limits_{-\infty}^\infty d\o\,  \E^{-\I \o(t-t_0)}
%\underbrace
{\sigma \E^{\l\o}
\over \E^{\b\bar \o}-1}%_{\makebox[0pt]{exponentially small 
%for $|\omega|\to\infty$,
%$\lambda\in(0,\beta)$}}
\varrho(k,\o) \stackrel{t_0\to-\infty}{\longrightarrow} 0 
\ee
for $\lambda\in(0,\beta)$ by Riemann-Lebesgue 
\cite{Landsman:1987uw}.\footnote{There are cases where
this line of reasoning breaks down. The decoupling
of the vertical part of the contour in RTF does however take place
provided the statistical distribution function in
the free RTF propagator defined below in (\ref{DcRTF1}) does
have as its argument $|k_0|$ and not the seemingly equivalent
$\sqrt{\vec k^2+m^2}$ \cite{Niegawa:1989dr,Gelis:1999nx}.}

With only $\mathcal C_1$ and $\mathcal C_2$ contributing, the
action in the path integral decomposes according to
\be\label{C12split}
\int\limits_{\mathcal C_1 \cup\, \mathcal C_2} \mathcal L(\ph)=
\int_{-\infty}^\infty dt\, \mathcal L(\ph^{(1)})-
\int_{-\infty}^\infty dt\, \mathcal L(\ph^{(2)})
\ee
where we have to distinguish between fields of type 1 (those
from contour $\mathcal C_1$) and of type 2 (those
from contour $\mathcal C_2$) because of the prescription
of \ind{contour ordering} in (\ref{greenfcts}).\footnote{Type-2
fields are sometimes called ``thermal ghosts'', which misleadingly
suggests that type-1 fields are physical and type-2 fields
unphysical. In fact, they differ only with respect to the
time-ordering prescriptions they give rise to.} From 
(\ref{C12split}) it follows that
type-1 fields have vertices only among themselves,
and the same holds true for the type-2 fields. However,
the two types of fields are coupled through the propagator,
which is a $2\times2$ matrix with non-vanishing off-diagonal
elements:
\be
{\bf D}^c(t,{ t'})=
\pmatrix{ \#{{\rm T} \ph(t){\ph(t')}} & {\s} 
\#{ {\ph(t')}\ph(t)} \cr
\#{ \ph(t){\ph(t')}} & \#{{\tilde{\rm T}} \ph(t){\ph(t')}} }.
\ee
Here $\tilde{\rm T}$ denotes anti-timeordering for the
2-2 propagator and $\sigma$ is a sign which is positive
for bosons and negative for fermions. 
The off-diagonal elements do not need a time-ordering
symbol because type-2 is by definition always later (on the contour) than
type-1.

In particular, for a massive scalar field one obtains
\bea\label{DcRTF1}
{\bf D}^c(k)&=&\pmatrix{ {\I\over k^2+m^2+\I \e} & 2\pi\delta^-(k^2-m^2) \cr
             2\pi\delta^+(k^2-m^2) & {-\I \over k^2+m^2-\I \e} } \nn\\
&&+
2\pi{\delta(k^2-m^2)}{ 1\over \E^{\beta |k_0|}-1} \pmatrix{1&1\cr1&1},
\eea
where $\delta^\pm(k^2-m^2)=\theta(\pm k_0)\delta(k^2-m^2)$.
The specifically thermal contribution is that of the second line.
Mathematically, it is a homogeneous Green function, as it should
be, because it is proportional to $\delta(k^2-m^2)$. Physically,
this part corresponds to Bose-Einstein-distributed, 
real particles on mass-shell.

The matrix propagator (\ref{DcRTF1}) can also be written
in a diagonalized form \cite{Matsumoto:1983gk,Matsumoto:1984au}
\be
{\bf D}^c(k)=\mathbf M(k_0) { \pmatrix{ {i G_F} & 0 \cr
             0 & {-\I G_F^*} } } \mathbf M(k_0)
\ee
with $G_F\equiv 1/(k^2+m^2+\I \e)$ and
\be
\mathbf M(k_0)= {1\over \sqrt{\E^{\beta|k_0|}-1}}
\pmatrix{\E^{{1\over2}\beta|k_0|} & \E^{-{1\over2}\beta k_0} \cr
         \E^{{1\over2}\beta k_0} & \E^{{1\over2}\beta|k_0|}}
\ee

In the
$T\to0$ limit one has
\be{\mathbf M(k_0)}
\stackrel{ \beta\to\infty}{\longrightarrow}\mathbf M_0(k_0)
=\pmatrix{1&\theta(-k_0)\cr
          \theta(k_0)&1}
\ee
so that one still has propagators connecting fields of
different type. However,
if all the external lines of a diagram are of the same type,
then also all the internal lines are,
because
$\prod_i \theta(k_{(i)}^0)=0$
when $\sum_i k_{(i)}^0 = 0$ and any connected region of
the other field-type leads to a factor of zero.

%%%%%%%% gt %%%%%%%%%%%

\section{Gauge Theories -- Feynman Rules}

As a simple application of the formalism developed above and
as a demonstration of the need for more formalism for gauge theories,
let us try to calculate the thermodynamic pressure of photons
in the imaginary-time formalism and in a covariant gauge.
\index{covariant gauges}
(There is no need for the real-time formalism here, because
we are not considering Green functions external lines.)
The simplest gauge to perform calculations is usually Feynman's
gauge, which simplifies the Lagrangian of the electromagnetic
fields according to
$\mathcal L = -\frac14 F_\mn F^\mn \to {1\over2} A_\nu \Box A^\nu$.

One would therefore expect the partition function to be
given by
$$\int_{\rm periodic} \!\!\!\!\!\!\!\!\!\!\!\! \mathcal D A \;
 \exp{\I \!\int\limits_0^{-\I {\beta}}\!dt\,d^3x\, \mathcal L} = 
{ \mbox{const.}\times}
(\det \Box)^{-{1\over2}\times 4}_{\rm periodic},
$$
and the thermal pressure would be calculated from
\bea
\ln Z &=& -4\times {1\over2} {\rm Tr} \ln \Box + \mbox{const} \nn\\
&=& -4\times {1\over2} V {\sum_n} 
\int\!{d^3k\over (2\pi)^3}\ln\({\omega^2_n}+k^2\)
{+ 
\;\mbox{const}} \nn\\
&=&4V \int\!{d^3k\over (2\pi)^3} \left[ -{{\beta} k\over 2} - 
\ln(1-\E^{-{\beta} k})
\right] {+ \;\mbox{const}} \nn
\eea
as
$$
P(T)-P(0)={1\over {\beta} V}\ln Z = 4T \int\!{d^3k\over (2\pi)^3} 
\ln(1-\E^{-{\beta} k})^{-1}
= { 2} \times  {\pi^2 { T^4}\over 45}
$$
giving {\em twice} the correct result of Planck for
blackbody radiation.

The error we made is that we have not calculated 
${\rm Tr} \E^{-{\beta} \bar H}$
in a physical Hilbert space (in fact in no Hilbert space at all, because
there are negative-norm states). Instead of only two physical (transverse)
degrees of freedom, we have added up the contributions from four.
The standard way to get rid of the unphysical degrees of freedom
is to cancel their contribution by ghost contributions, which evidently
are required already in the Abelian case.

\subsection{Path Integral -- Faddeev-Popov Trick}
\index{Faddeev-Popov Ghosts}

Because of gauge invariance under $\d A_\mu^a=D_\mu^{ab}(A)\omega^b$
(where $a,b$ are possible color indices), 
there is a redundancy
in the path integral that leads to zero modes in the kinetic
kernels, making them non-invertible. 
Thus, in order to be able to write down propagators
and do perturbation theory, one needs to remove this redundancy.
Using a suitable gauge fixing function $F^a[A](x)$, this 
can be done by inserting
\be\label{FPmeasure}
\prod\limits_{a,x} \delta\({ F^a[A](x)}\)
\cdot \det { \partial  { F^a} \over   \partial \omega^b}
\ee
into the measure of
the path integral, selecting only one gauge field configuration
per gauge orbit.

One can equally well use $F^a[A](x){-\zeta^a(x)}$ in place of
$F^a[A](x)$ with arbitrary functions $\zeta^a$ and perform
a Gaussian average 
$$\int\CD\zeta\, \E^{{\I\over 2\alpha}\zeta^2}\ldots$$
over the latter. This gives so-called
general or inhomogeneous gauge breaking terms 
\be\label{Lgf}
\CL\to\CL+{1\over 2\alpha}{ F^a[A]}^2
\ee
with gauge fixing parameter $\alpha$. 

In
\ind{covariant gauges} $F^a[A](x)=\partial ^\mu A_\mu^a(x)$
and Abelian electromagnetism, where $a$ takes only one value
and
$D_\mu^{ab}(A) \to \partial _\mu$,
the determinant in (\ref{FPmeasure})
is
$$\det {\partial F\over \partial \omega}=
\det\Box=(\det\Box)^{{+}{1\over2}\times 
{ 2}}$$
so this indeed compensates for the two unphysical degrees of freedom
in the above miscalculation of blackbody radiation.

Usually, this ``Faddeev-Popov determinant'' does not play a role in QED
because it is field independent. For calculating thermodynamic
potentials, it does, because this determinant
depends on temperature through the boundary conditions.

In non-Abelian gauge theories, 
$\det { \partial  F^a \over   \partial \omega^b}
=\det\({ \partial  F^a \over   \partial  A^b_\mu}D_\mu^{bc}
{ (A)}\)$ is
field dependent, and it is convenient to introduce
anticommuting and real\footnote{$\bar c$ is not
the conjugate of $c$, but an independent field.} Faddeev-Popov ghost fields
\be
\int \CD { c}\, \CD
{ \bar c}\, \exp \I \int_{ \CC} 
{ \bar c^a} { \partial  F^a \over   \partial  A^b_\mu}D_\mu^{bc}
{ (A)} { c^c} = \mbox{const.}\times \det\({ \partial  F^a \over
   \partial  A^b_\mu}D_\mu^{bc}
{ (A)}\).
\ee

The correct boundary conditions are clearly those of the
gauge potentials and thus are periodic in imaginary time
despite the fact that the ghosts are anti-commuting and thus
behave like fermions with regard to combinatorial factors
in front of Feynman diagrams \cite{Bernard:1974bq}.

%%%%%%%%%%%%%%%%%%%%%%%%%%%%%

\subsection{Covariant Operator Quantization}

\index{BRS quantization}
While the path integral makes it evident how to treat ghosts at
finite temperature, one can arrive at the same conclusion without
recourse to path integrals in covariant (BRS) operator quantization
\cite{Kugo:1978zq,Kugo:1979eg}, where at first it is somewhat
surprising that anticommuting fields should end up having
periodic rather than antiperiodic boundary conditions in
imaginary time.

\ind{BRS quantization} is preferably done with Lagrange multiplier fields $B$
and a  gauge-fixed Lagrangian (in general \ind{covariant gauges})
\be\label{LBRS}
\CL=\CL_{\rm inv}-A_\mu^a \partial ^\mu B^a + {\alpha\over 2}B^a B^a
-\I  (\partial ^\mu \bar c^a)
%(\underbrace{
D_\mu c
%}_{\makebox[0pt]{$\partial _\mu+gA_\mu\times c$}})^a
\ee
with $c$ and $\bar c$ being anticommuting field operators.

The gauge-fixed Lagrangian possesses a global fermionic (BRS) symmetry, 
which in any gauge reads \index{BRS transformation}
\be
\begin{array}{ll}
[\I Q_{\rm BRS},A_\mu]=D_\mu c, \qquad & \{\I Q_{\rm BRS},c\}
=-{g\over 2}c\times c, \\
{}[\I Q_{\rm BRS},B]=0, & \{\I Q_{\rm BRS},\bar c\}=iB,
\end{array}
\ee
where we used a vectorial notation to write for instance $D_\mu c=
(\partial _\mu+gA_\mu\times)c$.
In \ind{covariant gauges} for instance this is generated by
\be
Q_{\rm BRS}=\int\! d^3x \left[ 
B\cdot D_0c-c\cdot \partial _0B
+{\I g\over 2}(\partial _0\bar c)\cdot(c\times c)
\right].
\ee

The \ind{BRS operator} is nilpotent, $Q_{\rm BRS}^2=0$ and commutes with
Lagrangian and Hamiltonian, 
$[\I Q_{\rm BRS},\CL]=0=[\I Q_{\rm BRS},\CH]$.

There is one
further global symmetry, \ind{ghost number}, with conserved charge 
\be
N_c=\int\! d^3x \left[ \partial _0\bar c \cdot c - \bar c\cdot D_0 c \right]
\ee
satisfying
\be
[N_c,c]=c,\quad [N_c,\bar c]=-\bar c,\quad
[N_c,A_\mu]=0,\quad [N_c,B]=0.
\ee

$N_c$ is anti-hermitian, $N_c=-N_c^\dagger$, although it has 
real eigenvalues $n_{\rm gh}$, which is possible because
our arena is a non-Hilbert space $\CV$ containing negative-norm states.

The negative-norm states can be eliminated by the linear condition
\be
\CV \to \CV_{\rm phys}: Q_{\rm BRS}|{\rm phys}\rangle =0,
\ee
and the true physical Hilbert space is finally obtained by
modding out zero-norm states,
\be
{\CH_{\rm phys}}=\overline{{ \CV_{\rm phys}}/{
\CV_0}}\;.
\ee

The corresponding projection operator $\mathcal P$ in
$ \CH_{\rm phys}=\mathcal P  \CV$ can be shown \cite{Kugo:1978zq,Kugo:1979eg}
to have a complement that is ``BRS exact'', meaning
\be
{ \mathcal P}+{\{Q_{\rm BRS},
\mathcal R\}}=\mathbf 1,
\ee
but the actual construction of these operators is rather complicated.

However, we apparently need them to be able to define the trace restricted
to the physical Hilbert space that appears in
$Z={\rm Tr}\big|_{\CH_{\rm phys}}\!\!\!\E^{-\beta H}
={\rm Tr}\,{\CP}\,\E^{-\beta H}$
or in expectation values of observables.

%%%%%%%%%%%%%%%%%%%%%%%%%%%%%

\subsubsection{Hata-Kugo Trick} 
There is however an elegant trick that avoids the explicit
construction of $\cal P$ \cite{Hata:1980yr}:
\index{Hata-Kugo trick}
{}From
$[N_c,Q_{\rm BRS}]=Q_{\rm BRS}$ it follows that
%$N_c Q_{\rm BRS}=Q_{\rm BRS} (N_c+1)$\\
$N_c^n Q_{\rm BRS}=Q_{\rm BRS} (N_c+1)^n$ and therefore
$\E^{\I \pi N_c}Q_{\rm BRS}=Q_{\rm BRS} \E^{\I \pi(N_c+1)}=
-Q_{\rm BRS}\E^{\I \pi N_c}$, so
\be\label{eiNcQBRS}
\{\E^{\I \pi N_c},Q_{\rm BRS}\}=0.
\ee
This, together with {$N_c|\psi\rangle =0$} for 
$|\psi\rangle \in { \CH_{\rm phys}}$ 
can be used to write
\bea
Z&=&{\rm Tr}\,{ \CP}\,\E^{-\beta H}
={\rm Tr}\,{ \CP\,\E^{\I \pi N_c}}\E^{-\beta H} \nonumber\\
&=&{\rm Tr}\,{ \E^{\I \pi N_c}}\E^{-\beta H}
-\underbrace{
{\rm Tr}\,{\{Q_{\rm BRS},\mathcal R\}}\,\,
{ \E^{\I \pi N_c}}\E^{-\beta H}}_{\textstyle
{\rm Tr}\,\mathcal R{\underbrace{\{\E^{\I \pi N_c},Q_{\rm BRS}\}}_0} 
\E^{-\beta H}}
\eea
since $[Q_{\rm BRS},H]=0$.

So the comparatively simple operator ${ \E^{\I \pi N_c}}$ can
be used in place of the complicated $\CP$ to express $Z$,
and similarly thermal expectation values of gauge-invariant
observables $\CO$, through
a trace in unrestricted $\CV$ containing ghosts and other unphysical degrees
of freedom,
\be\label{HK}
Z={\rm Tr} \left[ \E^{\I \pi N_c} \E^{-\beta H} \right],
\quad \langle \CO \rangle =Z^{-1} {\rm Tr} \left[ \E^{\I \pi N_c} 
\E^{-\beta H} \CO\right].
\ee

This result shows that the anticommuting ghosts, which naturally
are subject to antiperiodic boundary conditions, acquire
a purely imaginary chemical potential 
\be\label{imchpot}
\mu_c=\I\pi/\beta.
\ee

In the ITF, the Matsubara frequencies of the ghosts are therefore
\index{Matsubara frequencies!ghost}
\be
z_\nu=2\pi \I{ (n-{1\over2})}/\beta+\mu_c=2\pi \I { n}/\beta, 
\quad n\in\mathbb Z
\ee
like those of
ordinary bosons. Thus, while they do have fermionic combinatorics
in Wick contractions and the like,
thermodynamically they behave like bosons.

In the RTF, where the matrix-valued propagator (\ref{DcRTF1})
can be written as\index{real-time formalism}
\bea
&&-\I  \mathbf D =\pmatrix{ G_F & 0 \cr
          0 & -G_F^* \cr}+(G_F-G_F^*)\times \nonumber\\
&&\quad\times{\sigma}
 {\pmatrix{\theta(\bar k_0)n(\bar k_0)+\theta(-\bar k_0)n(-\bar k_0) &
\mathrm{sgn}(\bar k_0)n(\bar k_0) \cr
\mathrm{sgn}(\bar k_0)({\sigma}+n(\bar k_0)) & 
\theta(\bar k_0)n(\bar k_0)+\theta(-\bar k_0)n(-\bar k_0) }}
\label{DcRTF2}\eea
with $\sigma=\pm$ for bosons/fermions,
$n(x)={1\over \E^{\b x}-{\sigma}}$, and $\bar k_0=k_0-\mu$,
we have $\sigma=-$ for the Faddeev-Popov ghosts, but
\be
 n_{\rm FD}(k_0-\I \pi/\beta) 
={1\over \E^{\beta k_0{-\I \pi}}+1}
={-} n_{\rm BE}(k_0),
\ee 
so the imaginary chemical potential (\ref{imchpot})
in effect negates $\sigma$ and replaces Fermi-Dirac by
Bose-Einstein statistics.

Only the signs arising in Wick contractions are those of
fermions, which shows that the Faddeev-Popov ghost propagators have
indeed the right form to be able to compensate for unphysical
degrees of freedom contained in the gauge boson propagator,
which naturally has $\sigma=+$ and Bose-Einstein statistical
factors. \index{Faddeev-Popov ghosts}

Compared to (\ref{DcRTF2}), the gauge boson propagator
also has a factor $\CG_\mn=(-g_\mn+(1-\a){k_\mu k_\nu\over k^2})$
in \ind{covariant gauges}.
We shall also consider other gauges in what follows, which
can be characterized by
\be\label{fgauge}
g_\mn\to \mathscr G_\mn=g_\mn-{k^\mu \tilde f^\nu + 
\tilde f^\mu k^\nu \over  \tilde f
\cdot k} + (\tilde f^2 - \a k^2){k^\mu k^\nu \over  (\tilde f
\cdot k)^2},
\ee
where $\tilde f$ is the momentum-space version of
$f$ in $F^a[A]=f^\mu A^a_\mu$.

Popular gauge choices besides the familiar \ind{covariant gauges}
include \ind{axial gauges} ($F^a=n^\mu A^a_\mu$, $n^\mu$ const.)
and \ind{Coulomb gauge}(s) ($F^a=\partial ^i A^a_i$).

In axial gauges, ghosts decouple completely, because the
Faddeev-Popov determinant $\det(n\cdot \partial )$ is field-
and temperature-independent, however they are fraught
with technical difficulties, already at zero temperature. 
The particularly attractive
``temporal'' gauge $n^\mu=\delta^\mu_0$, which does
not cause {\em additional} breaking of Lorentz symmetry,
is unfortunately inconsistent with periodic boundary conditions.
Relaxing those leads to rather complicated Feynman rules
in the ITF \cite{James:1990it}, while the RTF version seems more
tractable \cite{James:1990fd}, at least it does not appear to be
more problematic than at zero temperature.\index{temporal gauge}

\ind{Coulomb gauge} is in fact widely used at finite temperature, because
it also does not lead to additional Lorentz symmetry breaking.
However, it does have less simple Slavnov-Taylor identities
\cite{Dirks:1999uc} because ghosts do not decouple, although
they frequently do not contribute, since their (RTF) propagator
does not contain statistical distribution functions.

\subsection{Frozen Ghosts}
In \cite{Landshoff:1992ne}, alternative Feynman rules have
been proposed which avoid thermalized ghosts even in \ind{covariant gauges}.
In the RTF 
\index{real-time formalism}
one can %additionally 
switch off the interactions
as $t_0\to-\infty$, and define the physical Hilbert space in terms
of Abelianized in-states.
Without additional Lorentz symmetry breaking, physical in-states
can then be chosen as
\be |{\rm phys,in}\rangle =|{ A_{\rm phys.}}\mbox{-quanta}\rangle  
 |0\; { {\rm w.r.t.}\;
 A_{\rm unphys.},B,\bar c,c}\rangle 
\ee
with 
$A_{\rm phys.}^{a\mu}(k) = { \mathscr A^\mn(k)}  A^a_\n(k)$ and
\be\label{Amn}
\mathscr A^{0\mu}=0,\quad \mathscr A^{ij}=-(\d^{ij}-{k^ik^j\over \vec k^2}).
\ee

The 
unphysical
states correspond to
$$A_{\rm unphys.}^{a\mu}(k) = \(g^\mn-\mathscr A^\mn(k)\) A^a_\n(k),
\quad B^a,\quad \bar c^a,\; {\rm and}\; c^a.$$

The interaction-picture \index{interaction picture}
free Hamiltonian then separates into
two commuting parts
\be
H_{0I}={ H_{0I}^{\rm phys}}+{ H_{0I}^{\rm unphys}}, \quad
[{ H_{0I}^{\rm phys}},{ H_{0I}^{\rm unphys}}]=0, 
\ee
and, because
${ H_{0I}^{\rm unphys}}{|{\rm phys,in}\rangle =0}$,
thermal averages factorize:
\bea
&&{ \sum\limits_{\rm phys}
\langle {\rm phys,in}|}\E^{-\b H_{0I}} \cdots { A_{\rm phys.}}  \cdots
{ A_{\rm unphys.}} \cdots { \bar c} \cdots { c} \cdots 
{ |{\rm phys,in}\rangle }=\nonumber\\
&& \sum\limits_{\rm phys}
\langle {A_{\rm phys.}| \E^{-\b H_{0I}^{\rm phys.}}
\cdots A_{\rm phys.}  \cdots  |A_{\rm phys.}}\rangle 
\langle {0|{ \cdots A_{\rm unphys.} \cdots \bar c \cdots c \cdots }|0}\rangle  \nonumber
\eea
with the thermal Wick theorem applying to the first factor
under the latter sum, and the
$T=0$ Wick theorem to the second one.

%%%%%%%%%%%%%%%%%%%%%%%%%%%%%%%%

This leads to alternative Feynman rules for gauge theories in RTF
in which only the transverse projection of the gauge bosons
have thermal (matrix) propagators, and all other fields have 
only the $T=0$ limits of those. 
The rest of the Feynman rules is as usual in RTF. 
\index{real-time formalism}

E.g. in \ind{Feynman gauge} ($\a=1$)
the gauge boson and ghost propagators now read
\bea
\mathbf D_\mn &=& { -\I  \mathscr A_\mn\mathbf M
\pmatrix{ G_F & 0 \cr
          0 & -G_F^* \cr}\mathbf M} {
-\I (g_\mn-\mathscr A_\mn) \mathbf M_0 
\pmatrix{ G_F & 0 \cr
          0 & -G_F^* \cr}\mathbf M_0} \nonumber\\
&=&-g_\mn
\pmatrix{ {\I\over k^2+\I \e} & 2\pi\delta^-(k^2) \cr
             2\pi\delta^+(k^2) & {-\I \over k^2-\I \e} }
{\!-2\pi\delta(k^2)\mathscr A_\mn n(|k_0|)\pmatrix{1&1\cr1&1}},\;\;\\
\mathbf D^{\rm gh}&=&\pmatrix{ {\I\over k^2+\I \e} & 2\pi\delta^-(k^2) \cr
             2\pi\delta^+(k^2) & {-\I \over k^2-\I \e} }.\;
\eea
In general linear gauges one has to replace $g_\mn$ in the
vacuum part according to (\ref{fgauge}); the thermal part
remains unchanged.

Using these Feynman rules simplifies certain calculations
in covariant and other gauges \cite{Landshoff:1992ne},
because, although ghosts are present, they do not carry
statistical distribution functions, but are ``frozen''.
On the other hand, the usual cancellation of pinch singularities
in the RTF (absence of ``pathologies''), turns out be more
complicated, and occurs in general only upon Dyson resummations
\cite{Landshoff:1993ag}. \index{covariant gauges}

%%%%%%%%%%%%%%%%%%%%%%%%%%%%%%%%

\section{Gauge Dependence Identities}

As we have seen, perturbation theory and its Feynman rules
require the introduction of gauge fixing terms. Clearly,
physical results have to come out independent of those.
We shall therefore now study in detail to what extent
perturbative calculations will exhibit dependences on
the gauge fixing parameters by considering
\be\label{dFA}
F^a[A]\to F^a[A]+\delta F^a[A].
\ee
With $\delta F^a[A]\propto F^a[A]$, this also comprises
a possible variation of the gauge parameter $\alpha$ in
(\ref{Lgf}) or (\ref{LBRS}).

\subsection{Gauge Independence of the Partition Function}

\subsubsection{Path Integral}
The introduction of the gauge fixing term together with the
Faddeev-Popov determinant according to (\ref{FPmeasure})
was done in a way that picks one representative field configuration
from each gauge orbit.\footnote{At least perturbatively
this is guaranteed by the existence of $\partial  F/\partial \omega$;
non-perturbatively there may be obstructions to worry about.} 
So by construction the partition function
or averages of gauge-invariant operators are independent of
the gauge fixing terms. 
%Possible obstructions would be
%signalled by zeros of the Faddeev-Popov determinant, which
%would mean that the ghost propagator would not exist.

This can be checked explicitly by noting that the variation (\ref{dFA})
of the gauge fixing function can be written as
\be
\label{deltaF}
\delta F[A] =  { \partial  { F[A]} \over  \partial  A_\mu}
{D_\mu[A] 
\underbrace{\left[ {\partial  F[A] \over  \partial  A}
\cdot D[A] \right]^{-1} 
\delta F[A]}_{\textstyle \tilde\delta\xi[A]}}.
\ee
A corresponding change of the gauge breaking term ${1\over 2\alpha}(F^a)^2$
is thus equivalent to a gauge transformation 
$\tilde \delta A_\mu=D_\mu[A]\tilde\delta\xi$ with the above
non-local, field-dependent parameter $\tilde\delta\xi[A]$.

The invariant part of the action is of course invariant under
$A\to A+\tilde \delta A$, as are any gauge invariant operators that
might have been inserted, so it remains to check that the
path integral measure $\CD A$ together with the Faddeev-Popov determinant
is invariant, too. 
This can indeed be verified
by writing $\tilde\delta\CD A$ as 
${\rm tr}{\partial \tilde\delta\!A/\partial A}$,
and $\tilde\delta\det[{\partial  F \over  \partial  A} \cdot D]=
\det[{\partial  F \over  \partial  A}\cdot D]
\times \tilde\delta({\rm tr}\log[{\partial  F \over  \partial  A}\cdot D])
= \det[{\partial  F \over  \partial  A}\cdot D] 
\times {\rm tr}([{\partial  F \over  \partial  A}\cdot D]^{-1}
\tilde\delta[{\partial  F \over  \partial  A}\cdot D])$, and then using
that gauge transformations form a group \cite{Dewitt:1967ub,Kobes:1991dc}.

\subsubsection{Covariant Operator Formalism}

In the covariant operator formalism the Lagrangian in a general gauge $F$
can be written as
\be
\CL=\CL_{\rm inv} +B\cdot F+{\a\over 2}B\cdot B-\bar c\cdot [Q_{\rm BRS},F]
\ee
and variations of $F$ correspond to BRS transformations with
parameter $\bar c\cdot \delta F$: \index{BRS transformation}
\be
\delta\CL = B\cdot \delta F - \bar c \cdot [Q_{\rm BRS},
\delta F]=\{Q_{\rm BRS}, \bar c\cdot \delta F\}
\ee
It is always possible to construct an operator $\delta G$ such
that
$\delta H=\{Q_{\rm BRS}, \delta G\}$ 
(if no time derivatives are involved, one simply has
$\delta G=-\int\!d^3x\,\bar c\cdot \delta F$).

Gauge independence of the partition function and of thermal
averages of gauge invariant operators defined by (\ref{HK})
can be verified as follows \cite{Hata:1980yr}:

Variations of exponentiated operators can be expressed as
\be
\E^{A+\delta B}-\E^A=\int\limits_0^\lambda d\lambda\, \E^{\lambda A}
\,\delta B\, \E^{(1-\lambda)A} + O(\delta^2)
\ee
and using this one finds
\bea
\delta {\rm Tr} [\E^{-\beta H{ +\I \pi N_c}} {\CO}] 
&=&-{ \int_0^\beta d\lambda}\, 
{\rm Tr} [\E^{-{\lambda} H} {\{Q_{\rm BRS}, \delta G\}}
\E^{{\lambda} H}\E^{-\beta H}{ \E^{\I \pi N_c}} {\CO}] \nonumber\\
&=&-{ \int_0^\beta d\lambda}\, {\rm Tr} [\E^{-{\lambda} H}\,{ \delta G}\, 
\E^{{\lambda} H}
 \underbrace{ \{\E^{\I \pi N_c},Q_{\rm BRS}\} }_0 \E^{-\beta H}{\CO}] %= 0
\eea
because of cyclic invariance of the trace and 
%because of (\ref{eiNcQBRS}), since 
$[Q_{\rm BRS},\CO]=0$
for a gauge-invariant operator $\CO$.

%%%%%%%%%%%%%%%%%%%%%

\subsection{Gauge Dependence of Green Functions}

Green functions, i.e. thermal averages of products of field operators
like
$\langle \mathrm T_c \cdots A_\mu \cdots \psi \cdots 
\bar \psi \cdots \rangle $,
are, however, gauge-variant objects, and will therefore 
in general depend
on gauge fixing parameters.\footnote{Notice that gauge invariance and
gauge-fixing parameter independence are separate issues: a functional
of fields can be gauge invariant and yet depend parametrically
on the gauge-fixing function; conversely, independence of gauge
fixing parameters (within a class of gauges)
does not imply that a particular functional (e.g. of mean fields)
is a gauge-invariant one.}

In particular, the propagators of gauge and matter fields
will contain all sorts of gauge parameter dependences. Yet,
they are among the prime objects of linear response theory
as they are used to derive the properties of quasi-particles.

Historically, a stimulating failure was the attempt to extract
the damping constant of long-wavelength plasmons in a gluon
plasma from one-loop thermal perturbation theory. Some of
the results that were accumulated in the 80's are summarized
in Table \ref{Tabgamma}. These turned out to be gauge independent
in algebraic gauges but gauge-parameter dependent in
covariant ones. Moreover, in the latter the damping constant
came out with the wrong sign which some took as signal of
an instability of the perturbative ground state 
\cite{Hansson:1987un,Nadkarni:1988ti}. %,Parikh:1989ii}.
\index{plasmon!damping}

\begin{table}
\caption{Bare one-loop gluonic plasmon damping constant 
$\gamma(|{\vec k}|\to0)$
\index{plasmon!damping}}
\renewcommand{\arraystretch}{1.2}%{1.4}
\setlength\tabcolsep{5pt}
\begin{tabular}{ccr}
\hline\noalign{\smallskip}
$\gamma_{\rm pl.}/[{g^2TN\over 24\pi}]$ &  gauge & published \\
\noalign{\smallskip}
\hline
\noalign{\smallskip}
$-[{11\over 4}+({\a\over 2}-2)^2]$ &  \ind{covariant gauges} &
 %Kalashnikov \& Klimov 
1980 \cite{Kalashnikov:1980cy}\\
$+1$ &  \ind{temporal gauge} &  %Kajantie \& Kapusta 
1985 \cite{Kajantie:1985xx}\\
$+1$ &  \ind{Coulomb gauge} &  %Heinz et al 
1987 \cite{Heinz:1987kz}\\
$-[11+{1\over4}(1-\alpha)^2]$ &  background covariant gauges 
\cite{Abbott:1981hw}&  %Hansson \& Zahed 
1987 \cite{Hansson:1987un}\\
$-{45\over 4}$ &  { gauge-independent} effective action 
\cite{Vilkovisky:1984st,Rebhan:1987wp}  &  %Hansson \& Zahed 
1987 \cite{Hansson:1987un}\\
$-11$ &   gauge-independent pinch technique \cite{Cornwall:1985eu}&  %Nadkarni 
1988 \cite{Nadkarni:1988ti}\\
$\vdots$ & ~ & ~ \\
\hline
\end{tabular}
\label{Tabgamma}
\end{table}
%%%%%%%%%%%%%%%%%%%%%%%%%%%%%

It was in particular Pisarski \cite{Pisarski:1989vd} who argued
that these results were simply incomplete, and who together
with Braaten \cite{Braaten:1990kk,Braaten:1990mz,Braaten:1990it}
devised an appropriate resummation scheme. However, since
explicit calculations can only be performed in practice with
a rather limited choice of gauge parameters, it is important
to investigate more generally how gauge dependent the full
propagators are and whether they contain gauge-independent
information at all. To this end, we shall first derive
rather general ``gauge dependence identities''
\index{gauge dependence identities} and study their
consequences for the thermal Green functions of interest
\cite{Kobes:1990xf,%Rebhan:1990wc,
Kobes:1991dc}.

\subsubsection{Primary Diagrams}

In order to unclutter the relevant relations, we shall
temporarily switch to the compact notation of DeWitt 
\cite{Dewitt:1967ub}, where a single index
$i$ comprises
all discrete and continuous field labels (e.g. $i=(A,\mu,a,x)$)
and
Einstein's summation convention is extended to include integration
over all of space and time
(the latter along the contour $\CC$). 
This way, $\ph^i$ will represent an arbitrary
gauge or matter field
$\ph^i=\{A_{\mu_i}^{a_i}(x_i),\psi_{\sigma_i}^{a_i}(x_i),\ldots\}$;
only the Faddeev-Popov ghosts fields will be treated separately.

The generating functional of Green functions reads
\begin{equation}\label{ZJ}
Z[J]=\langle  \E^{\mathrm i J_i \ph^i} \rangle 
\quad\mbox{with}\quad
J_i \ph^i=\int\limits_\CC\! d^4x [J_{(A)\mu a}(x) A^{\mu a}(x)+\ldots]
\end{equation}
and depends implicitly on a gauge fixing functional $F^\alpha[\ph]$,
where $\alpha=(a_\alpha,x_\alpha)$ 
comprises both a group and a space-time index.

Information on the dependence on $F^\alpha$ can be obtained
either by using BRS techniques or equivalently by employing
the non-local gauge transformation of (\ref{deltaF}), which in
compact notation reads
\be
\delta\ph^i = D^i_\alpha[\ph] \delta\xi^\a \quad\mbox{with}\quad
\delta\xi^\alpha=\delta\xi^\alpha[\ph]=
-\CG^\alpha{}_\beta[\ph]\,\delta F^\beta[\ph]
\ee
where $D^i_\alpha$ is a generalized function containing
the gauge generators, and 
$\CG^\alpha{}_\beta[\ph]=-(F^\beta_{,i}D^i_\alpha)^{-1}[\ph]$
is the Faddeev-Popov ghost propagator in a background field $\ph$.
This immediately gives
\be
\label{deltaW}
\delta\ln Z[J]=\mathrm i
J_j \left\langle  D^j_\alpha[\ph]\CG^\alpha{}_\beta[\ph]\,
\delta F^\beta[\ph] \right\rangle [J]
\quad\mbox{under $F^\alpha\to F^\alpha+\delta F^\alpha.$}
\ee

\begin{figure}[t] %\sidecaption
{\includegraphics[width=0.55\textwidth]{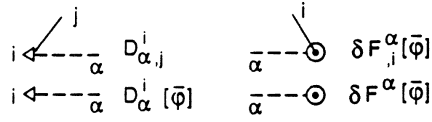}}
\caption{Additional Feynman rules for $\delta X^j[\bar \ph]$
in the case of linear gauge generators $D$ and
linear gauge fixing $F$}
\label{Figdxfr}
\end{figure}

\begin{figure}
\includegraphics[width=0.9\textwidth]{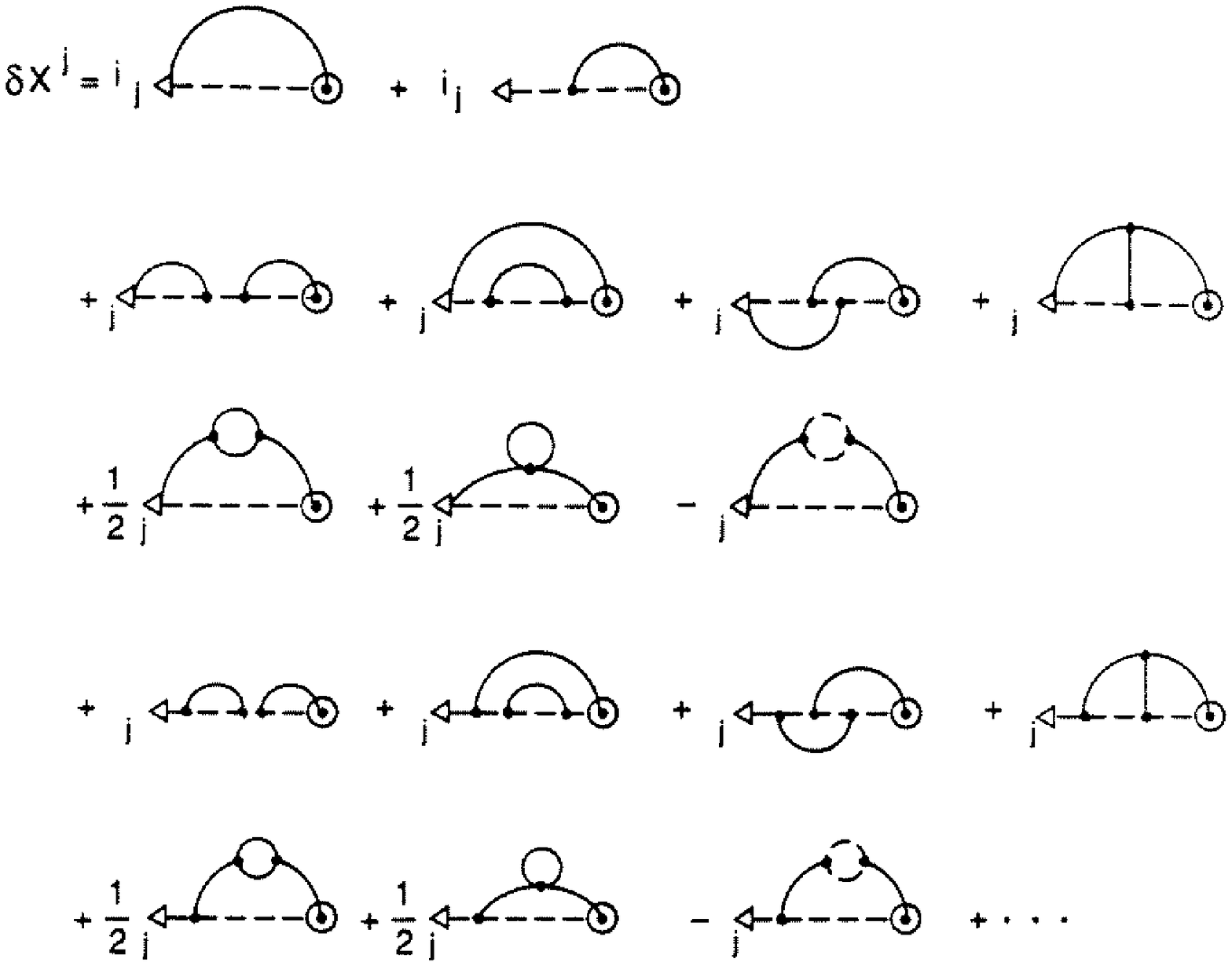}
\caption{Primary diagram expansion of $\delta X^j[\bar \ph]$
through 2-loop order. Contributions involving an undifferentiated
$\delta F^\alpha[\bar \ph]$ have been dropped, which corresponds
to omitting the trivial tree-level 
gauge dependence \protect\cite{Kobes:1991dc}}
\label{Figdxi}
\end{figure}

The diagrammatic content of (\ref{deltaW}) is more conveniently
investigated after a
Legendre transformation of $W[J]\equiv-i\ln Z[J]$
to the effective action
\be
\Gamma[\bar \ph]=W[J]-J_j \bar \ph^j,\qquad \bar \ph^j={\delta W[J]\over 
\delta J_j},
\ee
which is the generating functional of one-particle-irreducible (1-p-i)
diagrams. Equation (\ref{deltaW}) then becomes
\be
\delta\Gamma[\bar \ph]={\delta\Gamma[\bar \ph]\over \delta\bar \ph^j}
\left\langle  D^j_\alpha[\ph]\CG^\alpha{}_\beta[\ph]\,
\delta F^\beta[\ph] \right\rangle [\bar \ph]
\equiv \Gamma_{,j}[\bar \ph]\,  \delta X^j[\bar \ph] \;.
\ee
Diagrammatically, $\Gamma_{,j}[\bar \ph]$ is the sum of all
mean-field dependent (primary) 1-p-i one-point diagrams,
while $\delta X^j[\bar \ph]$ is given by primary diagrams
which involve the additional vertices introduced in Fig.~\ref{Figdxfr}
and which are 1-p-i except for the basic ghost line
attached to $\delta F^\beta$,
as shown in Fig.~\ref{Figdxi}.

\begin{figure}
\vspace*{20pt}
{\includegraphics[width=\textwidth]{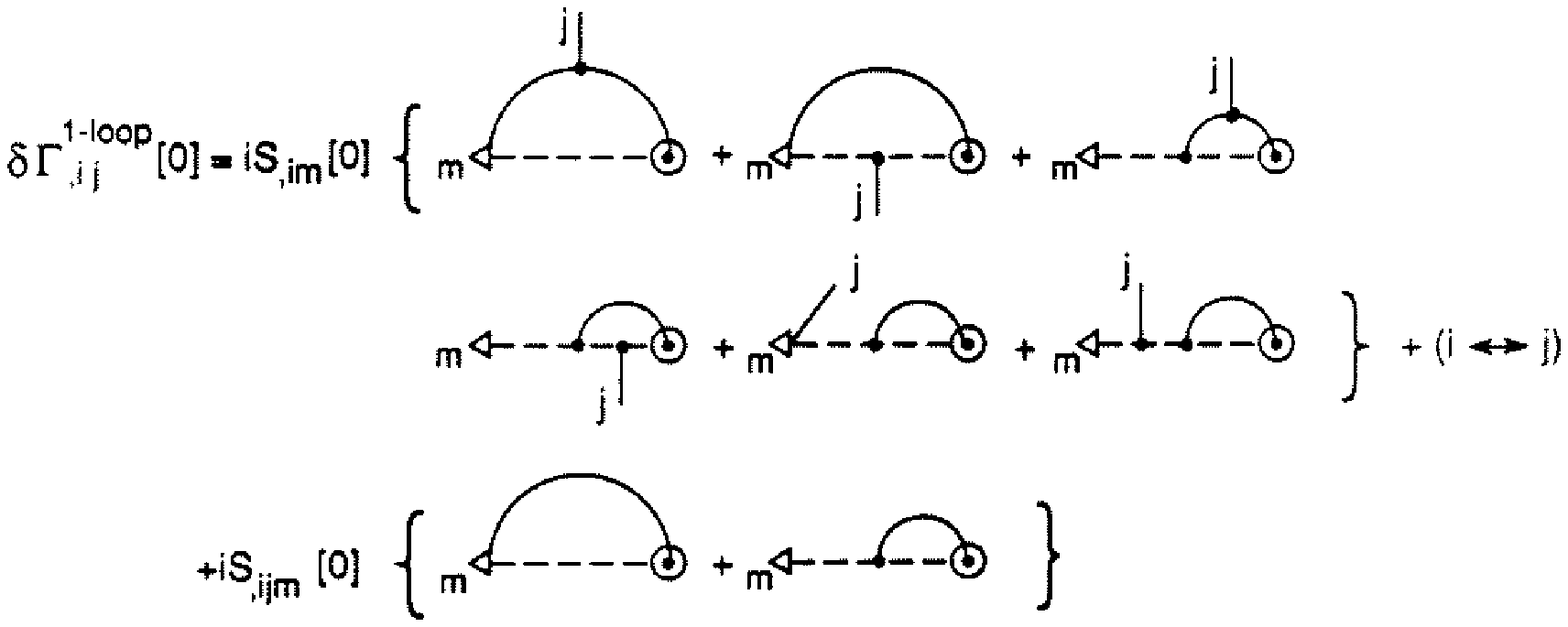}}
\caption{Gauge dependence of the 2-point vertex function
$\Gamma_{ij}[0]$ assuming that all one-point functions vanish 
at $\bar\ph=0$}
\label{FigdG2}
\end{figure}

{}From these relations one can derive gauge dependence
identities for 1-p-i vertex functions by differentiation
with respect to $\bar\ph$. For example, the 
gauge dependences of the 2-point
vertex function (self-energy) $\Gamma_{,ij}[0]$ are
determined by the diagrams shown in Fig.~\ref{FigdG2}.

\subsubsection{QED}

As a first application let us consider the case of an Abelian
theory such as QED. The additional Feynman rules of
Fig.~\ref{Figdxfr} involve the ghost propagator,
but there are no further ghost vertices in the theory
(for linear gauge fixing), so only the very first diagram
in Fig.~\ref{FigdG2} arises.

Furthermore,
the structure of the gauge generator is such that
\be
\raisebox{-5pt}[0pt][0pt]{\includegraphics[width=0.09\textwidth]{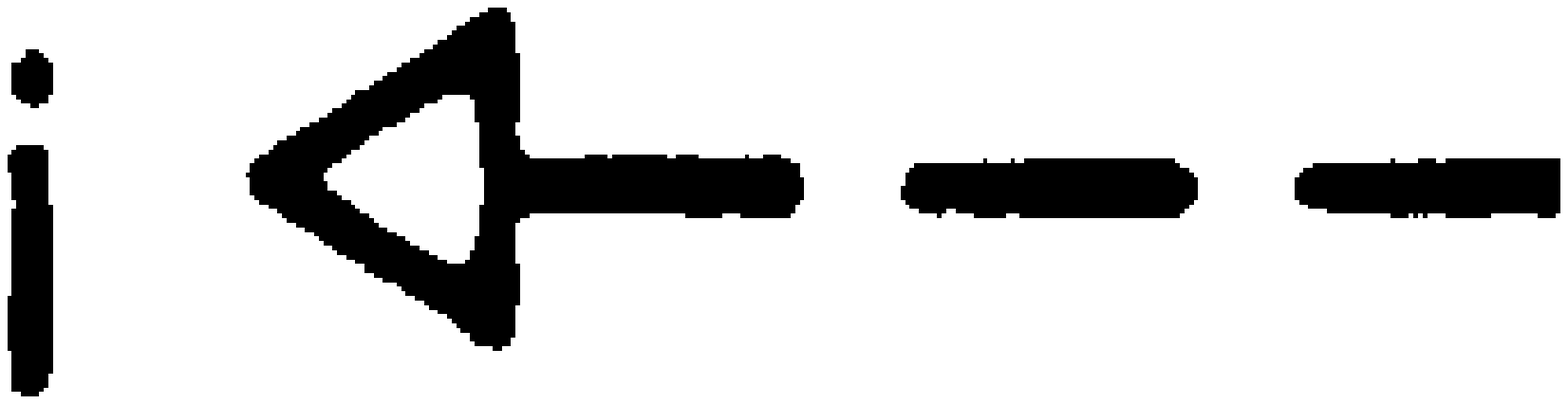}}
= \left\{ \begin{array}{ll}
\partial _{\mu_i}^{(x_i)} & \mbox{if $i \leftrightarrow A_{\mu_i}(x_i)$} \\
0 & \mbox{else} \end{array} \right. \;
\raisebox{-9pt}[0pt][0pt]{\includegraphics[width=0.09\textwidth]{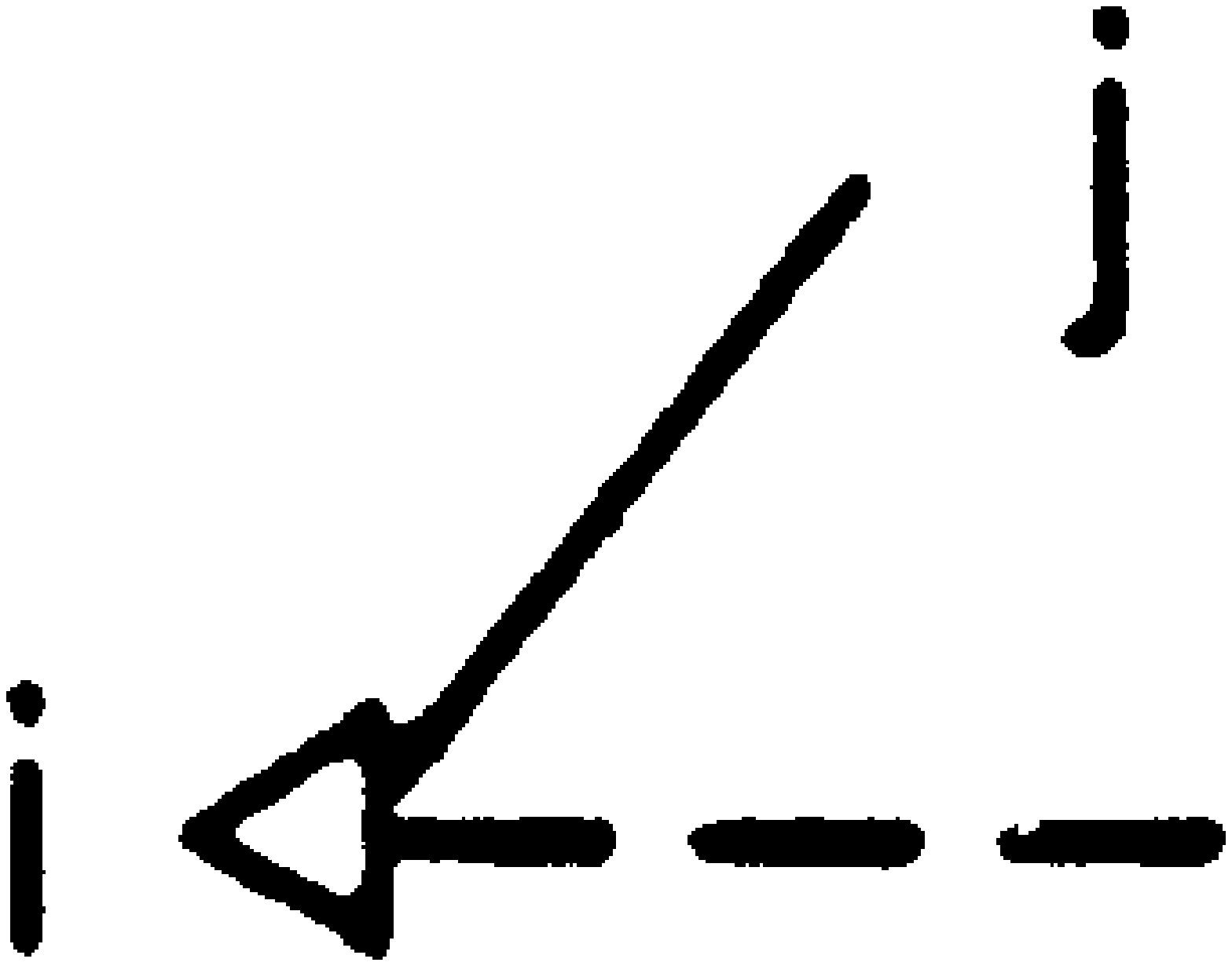}}
= \left\{ \begin{array}{ll}
0 & \mbox{if $i,j \leftrightarrow A_\mu$} \\
\pm \mathrm i e & \mbox{if $i,j \leftrightarrow \stackrel{(-)}{\psi}$}
  \end{array} \right.
\ee

If the external indices $i,j$ of $\delta\Gamma_{,ij}$
correspond to photons, one finds that one cannot even
build the one remaining diagram of Fig.~\ref{FigdG2},
so $\delta\Gamma_{A^\mu A^\nu}$ proves to be completely
gauge-fixing independent. This is in fact a well-known
result which can be understood also by the gauge invariance
of the electromagnetic current operator.

\begin{figure}
\includegraphics[width=0.7\textwidth]{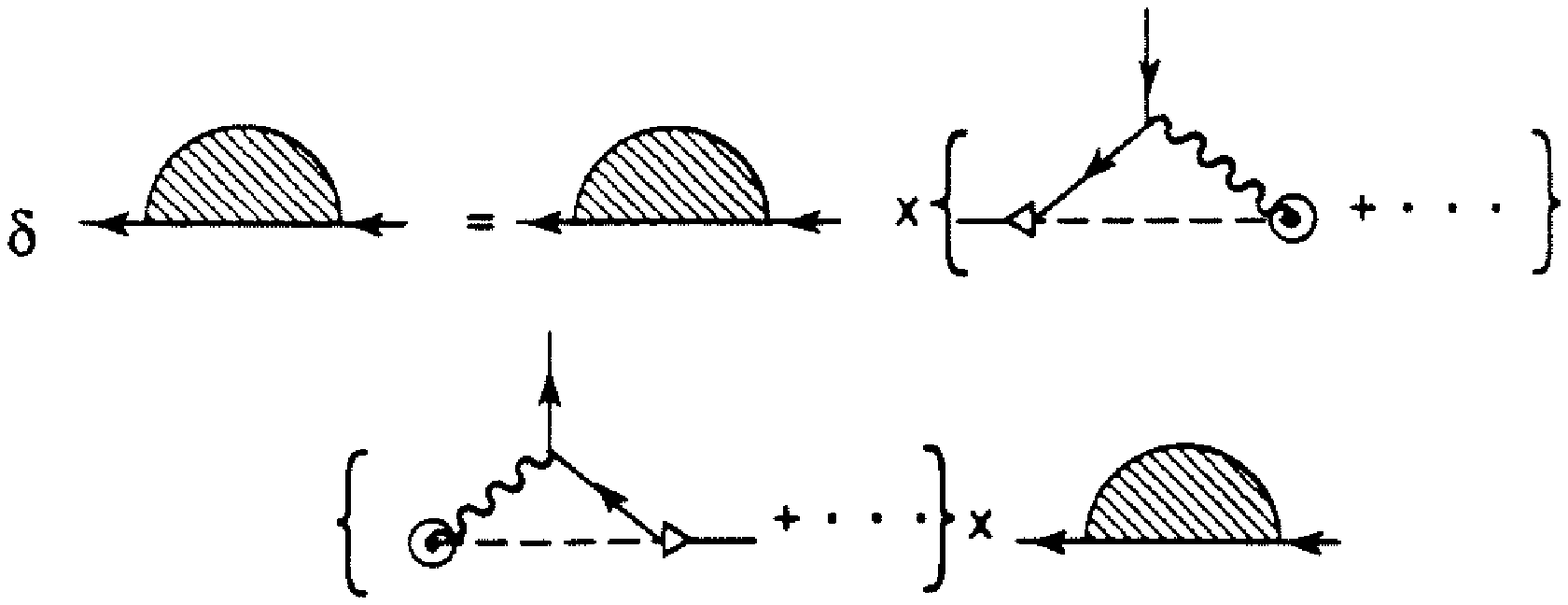}
\caption{Gauge dependence identity for the fermion self-energy}
\label{FigdSigma}
\end{figure}

On the other hand, if the external lines are fermionic,
there is a non-trivial right-hand-side to Fig.~\ref{FigdG2},
as shown in Fig.~\ref{FigdSigma}, so the fermion self-energy
is a gauge fixing dependent quantity, already in QED.

\subsubsection{Hard Thermal/Dense Loops}\index{HTL}\label{giHTL}

In the high-temperature (large-chemical-potential) limit of
QED and QCD, it turns out that the leading contributions
to the 1-loop vertex functions, the so-called ``hard thermal
(dense) loops'' (HTL/HDL) obey tree-level-type ghost-free 
\ind{Ward identities} and appear to be gauge-fixing independent
\cite{Frenkel:1990br,Braaten:1990mz,Braaten:1990az}.
This gauge independence, however, does not arise in an
obvious way and involves non-trivial cancellations
in the various gauges that have been considered.

The above gauge dependence identities can be used to
verify the gauge independence of the HTL's in a rather
simple manner. The only further ingredients needed are
the \ind{temperature power-counting} rules given in 
\cite{Braaten:1990mz}, which, roughly, read as follows:
in a Feynman diagram, explicit loop momenta in the numerator
give a factor $T$, each propagator counts as $T^{-1}$,
and the sum-integral over the loop momentum contributes
$T^3$ unless there are two or more propagators with
the same statistics, in which case the sum-integral
counts as $T^2$.

By this, the leading temperature contributions to a 1-loop
vertex function are found to be proportional to $T^2$,
such that an $N$-point gluon vertex function scales as
$\Gamma_{,(N)} \sim g^N T^2 k^{2-N}$ (where $k$ represents
generically components of external momenta). If two external
lines are fermionic, we have $\Gamma_{,(N)} \sim g^N T^2 k^{1-N}$,
while
vertex functions with more than two external fermion lines
do not contribute terms $\propto T^2$.

Considering e.g. vertex functions with only external gluons,
all of the potential gauge dependences of the HTL's are
contained in the 1-loop contributions to 
$\delta\Gamma_{,(N)} = \sum\limits_{M=0 \atop perms.}^N
\Gamma_{,(N+1-M)} \delta X^\cdot_{,(M)}$ with
the diagrammatic structure of
$\delta X^\cdot_{,(M)}$
as given by
\be
\label{dxiN}
\delta X^\cdot_{,(M)}=\sum_{\rm perm.} \biggl\{ \;\;
\raisebox{-38pt}{\includegraphics[width=0.5\textwidth]{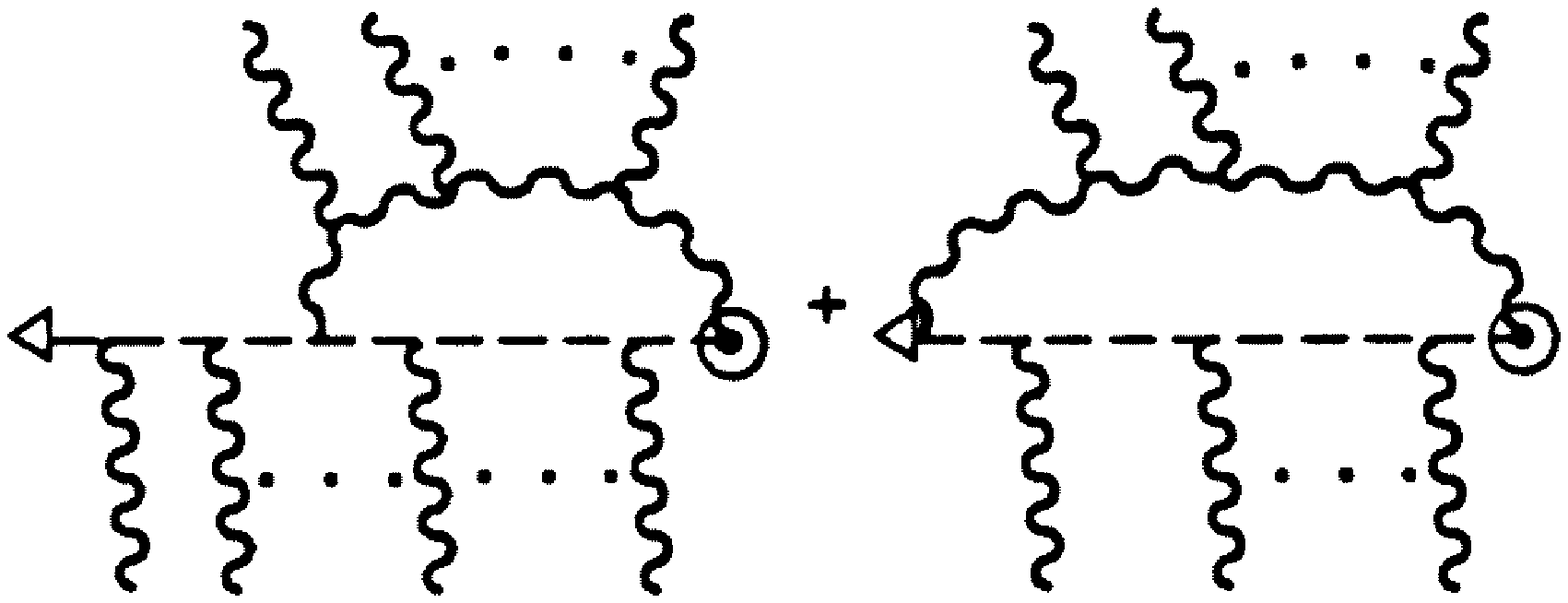}}
\biggr\}
\ee

The above \ind{temperature power-counting} rules are modified only
by the additional vertex $\delta F^\alpha_{,i}[0]$ which
may bring in one power of loop momentum and thus one power of $T$
in derivative gauges, or none in algebraic ones. Adding up,
one finds that the right-hand-side of (\ref{dxiN}) is
proportional to $T^{0\ldots 1}$. Hence, gauge dependences
of one-loop vertex functions can occur only at subleading order
$\propto T$. %; within classes of algebraic gauges even
%only $\propto T^0$.
HTL(HDL)'s on the other hand are found to be completely gauge-fixing
independent.

\subsection{Gauge Independence of Propagator Singularities}

In non-Abelian gauge theories,
all the matter and gauge field vertex functions, self-energies,
and propagators
contain highly nontrivial gauge
dependences, which raises the question whether there 
is any gauge-independent and therefore potentially physical
information in those at all.
We shall give an affirmative answer by
showing that (the locations of)
certain singularities of
the full propagators are indeed gauge independent %, 
%and so are all their coefficients in a {\em systematic} approximation scheme
\cite{Kobes:1990xf}.

\subsubsection{Non-Abelian Gauge-Boson Propagator}
We begin by analysing the
Lor\-entz structure of the gluon propagator in the case
of a general gauge that preserves the rotational symmetry.
Moreover, we shall simplify things by dropping
any color indices, which presupposes
the absence of color symmetry breaking.\footnote{The extension
of the following results to color superconducting situations
has not yet been worked out, but would be of great
interest in view of the gauge dependence issues
there \cite{Rajagopal:2000rs}.}
\index{gluon propagator}

In momentum space one can define a transverse projection
of the 4-velocity of the heat bath, 
$\tilde n^\mu = (g^{\mu\s}-{k^\mu k^\s \over  k^2})u_\s$, and use it to
write the general structure of the gauge-boson propagator as
\be
G^\mn(k)=\Delta_A \mathscr A^\mn + \Delta_B \mathscr B^\mn + 
\Delta_C \mathscr C^\mn + \Delta_D \mathscr D^\mn
\ee
with 
\be\label{ABCD}
\begin{array}{l@{\qquad}l}
\displaystyle\mathscr A^\mn(k) =
[g^\mn - {k^\mu k^\nu \over  k^2}] - 
{\tilde n^\mu \tilde n^\nu \over  \tilde n^2}\,,
&\displaystyle
\mathscr B^\mn(k) = 
{\tilde n^\mu \tilde n^\nu \over  \tilde n^2} \,,\\[8pt]
\displaystyle\mathscr C^\mn(k) = {1\over |{\vec k}|} \left\{ 
\tilde n^\mu k^\nu + k^\mu \tilde n^\nu \right\}\,, 
&\displaystyle\mathscr D^\mn(k) = {k^\mu k^\nu \over  k^2}\,. 
\end{array}
\ee

Here $\mathscr A^\mn$ is the spatially transverse projector
introduced already in (\ref{Amn}), and $\mathscr B^\mn$ is a second,
independent tensor that is likewise transverse with respect to 4-momentum,
but longitudinal with respect to 3-momentum.
$\mathscr C^\mn$ and $\mathscr D^\mn$ complete the basis of
symmetric tensors, with $\mathscr C^\mn$ chosen such that
$k_\mu \mathscr C^\mn k_\nu=0$, and $\mathscr D^\mn$ longitudinal
with respect to 4-momentum.

$\mathscr A$, $\mathscr B$, and $\mathscr D$ are idempotent,
whereas $\mathscr C^2=-(\mathscr B+\mathscr D)$. 
Under a Lorentz trace, products of one such tensor with a different
one vanish; without trace, $\mathscr A$ is orthogonal to
all the others, but among the rest one only has $\mathscr B\perp D$.

Similarly, we shall decompose the
gluon self-energy 
$\Pi^\mn = {G}^{-1\mn}-{G}^{-1\mn}_0$ according to
\be
\Pi^\mn = -\Pi_A \mathscr A^\mn - \Pi_B \mathscr B^\mn 
- \Pi_C \mathscr C^\mn - \Pi_D \mathscr D^\mn\;.
\ee

%%%%%%%%%%%%%%%

%\subsubsection{Hard thermal$|$dense loop photon/gluon propagator}
\index{HTL!gluon propagator}
At %energy/
momentum 
scales $\omega,k \ll T,\mu$, the leading-order
term in the one-loop polarisation tensor $\Pi^\mn$ is given by the HTL (HDL)
$\sim {\rm max}(T^2,\mu^2)$, which has only 4-d-transverse contributions
\begin{subeqnarray}
\Pi_A^{\rm HTL} &=& {1\over2} (\Pi^{\rm HTL}{}_\mu{}{}^\mu - \Pi^{\rm HTL}_B)\\
\Pi_B^{\rm HTL} &=& -{k^2\over {\vec k}^2} \Pi^{\rm HTL}_{00}\label{PiBHTL}\\
\Pi_C^{\rm HTL} &=&0\\
\Pi_D^{\rm HTL} &=&0
\end{subeqnarray}
where, at high $T$,
\bea
&&\Pi^{\rm HTL}{}_\mu{}{}^\mu={{ e^2T^2}\over 3},\label{PiHTLmumu}\\
&&\Pi^{\rm HTL}_{00}={{ e^2T^2}\over 3}
\(1-{k^0\over 2|{\vec k}|}\ln{k^0+|{\vec k}|\over k^0-|{\vec k}|}\). 
\label{PiHTL00}
\eea
As a function of frequency and 3-momentum, the result is
identical in QED and QCD, if for the latter we define
$e^2 := g^2(N+N_f/2)$ (for ${\rm SU}(N)$ with $N_f$ flavors)
\cite{Kalashnikov:1980cy,Weldon:1982aq}.
If there is also a nonvanishing chemical potential $\mu$, 
a similar result holds where $T^2\to T^2+{3\over \pi^2}\mu^2$ 
in terms $\propto N_f$ (all of them
in QED)---with obvious generalization to the case
of different chemical potentials $\mu_i$ for different flavors $i$.

The HTL-dressed propagator $G^{\rm HTL}=(G_0^{-1}+\Pi^{\rm HTL})^{-1}$
has poles off the usual light-cone, which come in two branches
determined by
\begin{subeqnarray}
\Delta_A^{\rm HTL-1}&=&k^2 - \Pi_A^{\rm HTL}=0\label{DAHTL}\\
\Delta_B^{\rm HTL-1}&=&k^2 - \Pi_B^{\rm HTL}=0\label{DBHTL}
\end{subeqnarray}
Since, as we have seen above, the HTL contribution is
completely gauge independent and the gauge fixing
parameters contained in $G_0^{-1}$ do not 
appear in (\ref{DAHTL},\ref{DBHTL}), 
the $A$- as well as the $B$-part of the HTL propagator is
completely gauge independent.

\begin{figure}
%\centerline{
\includegraphics[viewport = 0 160 540 550,scale=0.44]{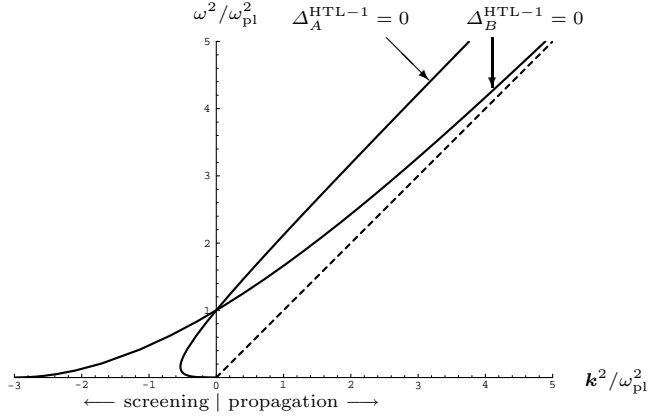} %}

\bp\scriptsize
%\put(20,120){\fbox{\footnotesize quadratic scales!}}
\put(218,18){$\vec k^2/\omega_{\rm pl}^2$}
\put(70,157){$\omega^2/\omega_{\rm pl}^2$}
%\put(140,160){transv. gluon}
\put(107,155){$\Delta_A^{{\rm HTL}-1}=0$}
%\put(180,160){long. plasmon}
\put(173,155){$\Delta_B^{{\rm HTL}-1}=0$}
\put(143,150){\vector(1,-1){16}}
\put(183,150){\vector(0,-1){19}}
\put(78,10){$|$ propagation $\longrightarrow$}
\put(28,10){$\longleftarrow$ screening}
\ep
\caption{The location of the zeros of $\Delta_A^{{\rm HTL}-1}$ (transverse
gluons) and of $\Delta_B^{{\rm HTL}-1}$ (longitudinal plasmons) in quadratic
scales such as to show propagating modes and screening phenomena
on one plot. Above a common \ind{plasma frequency} $\omega_{\rm pl.}$
there are propagating quasi-particle modes, which for large
momenta in branch A
tend to a mass hyperboloid with asymptotic mass 
$m_\infty^2={3\over 2}\omega_{\rm pl.}^2$, and in branch B approach the
light-cone exponentially with exponentially vanishing residue.
For $\omega<\omega_{\rm pl.}$, $|\mathbf k|$ is the inverse screening
length, which in the static limit vanishes for mode A (absence of
magnetostatic screening), but reaches the \ind{Debye mass},
$\hat m_D^2=3\omega_{\rm pl.}^2$, for mode B
(electrostatic screening)
\index{plasmon}
\index{asymptotic thermal mass}
\label{Figg}}
\end{figure}

The physical interpretation of the $A$- and $B$-branch of
propagator poles (displayed in Fig.~\ref{Figg})
is that the former represents quasi-particles
which are in-medium versions of the physical polarisation
of the gauge bosons, while the appearance of the $B$-branch is a purely
collective phenomenon corresponding to charge density oscillations
(plasmons) above the plasma frequency and to charge screening below.
\index{plasmon}
%The residues of these latter poles turn out to disappear exponentially
%for momentum scales $\gg eT$ so that this branch is relevant
%only at soft scales $\sim eT$. 

%%%%%%%%%%%%%%%%%%%%%%%%

%{\bf Beyond LO}:

Beyond the HTL approximation and in non-Abelian theories, however, 
one has gauge parameter dependences within $\Pi$, and
also $\Pi_\mn k^\mu\not=0$ so that
$\Pi_C\not=0, \Pi_D\not=0$.

Considering a general, rotationally invariant gauge
$\tilde f^\mu(k) \tilde A_\mu(k)$ as in (\ref{fgauge}), this
can be parameterized as
\be
\tilde f^\mu(k)=
\tilde\beta(k) k^\mu+\tilde\gamma(k)\tilde n^\mu.
\ee
Covariant gauges then correspond to $\tilde\beta=1,\tilde\gamma=0$,
Coulomb gauges to $\tilde\beta=\tilde n^2,
\tilde\gamma=-k^0$, and temporal axial gauge to 
$\tilde\beta=k^0/k^2, \tilde\gamma=1$.
\index{covariant gauges}
\index{Coulomb gauge}
\index{temporal gauge}

%BRS $\Rightarrow
%\tilde f_\mu \tilde f_\nu {G}^\mn = \alpha$ \\
%$\then$ only: $ \Pi_D \(k^2-\Pi_B\) = \Pi_C^2$\\
%%\rightline{\footnotesize $\Rightarrow$ $ \Pi_D=O$(2-loop)}

The structure functions of the gauge propagator become more
complicated, to wit,
\begin{subeqnarray}
\Delta_A &=& [ k^2 - \Pi_A ]^{-1} \\
\Delta_B &=& [ k^2 - \Pi_B -
{2\tilde \b\tilde \g |{\vec k}| \Pi_C 
- \a \Pi_C^2 + \tilde \g^2 {\tilde n}^2 \Pi
_D  \over  \tilde \b^2 k^2 - \a \Pi_D } ]^{-1} \\
\Delta_C &=& -{\tilde \b\tilde \g |{\vec k}| - \a \Pi_C \over  
\tilde \b^2 k^2 - \a \Pi_D} \\
\Delta_D &=&
{ \tilde \g^2 \tilde n^2 + \a(k^2-\Pi_B) \over  \tilde \b^2 k^2 - \a \Pi_D }
 \Delta_B \label{DeltaD}
\end{subeqnarray}
and there are gauge parameters everywhere, both explicitly
and also within the structure functions of $\Pi$.

%%%%%%%

These gauge dependences are controlled by the
\ind{gauge dependence identities}
discussed above, and, in compact notation, they have the form
\be
\delta\Delta^{ij}\Big|_{J=0}=
-(\Delta^{im}\delta X^j_{,m}+\delta X^i_{,m}\Delta^{mj})\Big|_{J=0}
\ee
for the full propagator.
Specialized to the thermal gauge-boson propagator in $f^\mu$-gauge,
one finds \cite{Kobes:1990xf,Kobes:1991dc}
\begin{subeqnarray}
\delta{ \Delta_A^{-1}}&=&{ \Delta_A^{-1}}\Bigl[-\mathscr A^\mu_\nu(k)\delta
X^\nu_{,\mu}(k)\Bigr] \equiv { \Delta_A^{-1}}\delta Y \label{dDA} \\
\delta{ \Delta_B^{-1}} &=& { \Delta_B^{-1}}
\left[-{\tilde n^\mu\over \tilde n^2}+
{\tilde\gamma\tilde\beta-\alpha \Pi_C/|\vec k| \over 
\tilde\beta^2 k^2 -\alpha \Pi_D}k^\mu\right]2\tilde n_\nu \delta X^\nu_{,\mu}
\equiv { \Delta_B^{-1}}\delta Z \label{dDB} 
\end{subeqnarray}
but no such relations for $\Delta_C$ and $\Delta_D$.

If $\delta Y$ and $\delta Z$ are regular on the two ``mass-shells''
defined by $\Delta_A^{-1}=0$ and $\Delta_B^{-1}=0$, the
relations (\ref{dDA},\ref{dDB}) imply that the locations of these particular
singularities of the gluon propagator are gauge fixing independent,
for if $\Delta_A^{-1}=0=\Delta_B^{-1}$ then also
$\Delta_A^{-1}+\delta\Delta_A^{-1}=0=\Delta_B^{-1}+\delta\Delta_B^{-1}$.
%On the other hand, residues (if those singularities are simple poles
%at all) are not protected and may be gauge dependent; even the
%nature of the singularity may be different from gauge to gauge!

So everything depends on whether 
the possible singularities of $\delta X^\nu_{,\mu}$
could coincide with the expectedly physical dispersion laws
$\Delta_A^{-1}=0$ and $\Delta_B^{-1}=0$. Diagrammatically,
$\delta X^\nu_{,\mu}$ is obtained from the primary diagrams
of Fig.~\ref{Figdxi} by inserting one additional vertex in
all possible ways (and omitting all resulting tadpole-like diagrams
in the case of no spontaneous symmetry breaking). Since
the primary diagrams are 1-particle reducible with respect
to the basic ghost line attached to $\delta F^\alpha_i$,
$\delta X^\nu_{,\mu}$ will have singularities like the (full)
ghost propagator. These singularities are however generically different
from those that define the spatially transverse and longitudinal
gauge-boson quasi-particles. Indeed, in leading-order thermal
perturbation theory, the \ind{temperature power-counting} rules
referred to in Sect.~\ref{giHTL} imply that
the ghost propagator does not receive contributions $\sim e^2 T^2$
and thus will have completely different dispersion laws.

The other parts of the diagrams making up $\delta X^\nu_{,\mu}$ are 1-p-i
and may develop singularities for other reasons, namely when
one line of such an 1-p-i diagram is of the same type as the
external one, and the remaining ones are massless. This may potentially
give rise to infrared or mass-shell singularities.
However, these singularities will be absent as soon as
an overall infrared cut-off is introduced, for example
by considering first a finite volume. In every finite volume,
this obstruction to the gauge-independence proof is then
avoided, and $\Delta_A^{-1}=0$ and $\Delta_B^{-1}=0$ define
gauge-independent dispersion laws if the infinite volume limit
is taken last of all \cite{Rebhan:1992ak}.

This reasoning leads to the conclusion that the positions
of all the singularities
of $\Delta_A$ are gauge-fixing independent,
though not necessarily their type or e.g.\ their
residues if they are simple poles. In the case
of $\Delta_B$, there is a slight complication by the
contents of the square bracket in (\ref{dDB}). There is
a kinematical pole $1/k^2$ hidden in the $\tilde n$'s, and
there is a contribution from the obviously gauge-dependent
$\Delta_D$ (cf.\ (\ref{DeltaD})). These gauge artefacts
have to be excluded, but they are gauge dependent already
at tree level and thus easy to identify. For example, $\Delta_B$
as defined above has a factor of $k^2$ which cancels in the
\ind{Coulomb gauge} propagator but not in that of \ind{covariant gauges},
so this massless mode is a gauge mode and thus unphysical.

The gauge-(in)dependence identities (\ref{dDA},\ref{dDB}) also explain
the gauge dependences found in the one-loop calculation
of the plasmon damping constant mentioned above. \index{plasmon!damping}
Truncating e.g. (\ref{dDA}) at one-loop order
gives
\be\label{dDA1}
\delta{ \Delta_A^{-1}{}^{(1)}}(k)={ \Delta_A^{-1}{}^{(0)}}\delta Y^{(1)},
\ee
with superscripts referring to bare loop order and using
that $\delta Y^{(0)}\equiv0$. However, the HTL plasma dispersion
law is derived from $\Delta_A^{-1}{}^{(0)} + \Delta_A^{-1}{}^{(1)}=0$,
and the ``correction'' $\Delta_A^{-1}{}^{(1)} \sim \Pi^{\rm HTL}
\sim g^2 T^2 \sim \omega_{\rm pl.}^2$ is not small but
sets the scale for everything. The temperature-power-counting rules
of Sect.~\ref{giHTL} give $\delta Y^{(1)} \sim g^2 T/\omega_{\rm pl.}$, so the
right-hand side of (\ref{dDA1}) does not vanish at the order of
the damping contribution $\gamma\times\omega_{\rm pl.} 
\sim g\omega_{\rm pl.}^2$.

On the other hand, if one does have a good expansion parameter
(which bare loop order obviously
is not), then the identities (\ref{dDA},\ref{dDB})
imply order-by-order gauge independence. %, provided all the prerequisites
%discussed above are fulfilled. 

As will be discussed further below, HTL perturbation theory 
\cite{Braaten:1990kk,Braaten:1990mz}
claims
to be a systematic framework, although not up to arbitrarily high
orders, and the expansion parameter is essentially $\sqrt{g^2}$.
In \cite{Braaten:1990it}, the long-wavelength plasmon damping
\index{plasmon!damping}
constant has been calculated by Braaten and Pisarski with the result
$\gamma(|\vec k|=0)/[{g^2TN\over 24\pi}] = + 6.635\ldots$
and formal checks as to its gauge independence were positive.

More explicit calculations by Baier et al., however, revealed
that, in \ind{covariant gauges} 
and \index{plasmon!damping}
on plasmon-mass-shell, HTL-resummed perturbation theory
still leads to explicit gauge dependent
contributions to the damping of fermionic \cite{Baier:1992dy}
as well as gluonic \cite{Baier:1992mg}
quasi-particles. But, as was pointed out in \cite{Rebhan:1992ak},
these apparent gauge dependences are avoided if the quasi-particle
mass-shell is approached with a general infrared cut-off such
as finite volume, and this cut-off lifted only in the end.
This procedure defines gauge-independent dispersion laws and
the gauge dependent parts are found to pertain to the residue,
which at finite temperature happens to be linearly infrared
singular in \ind{covariant gauges}, rather than only logarithmically
as at zero temperature, due to Bose enhancement.

\subsubsection{Extension to Fermions}

The fermion propagator at non-zero temperature or density
has one more structure function than usually. In the
ultrarelativistic limit where masses can be neglected,
the fermion self-energy can be parametrized according to
\be\Sigma(k_0, {\vec  k})\,=\,a(k_0, k)\,\gamma^0\,+\,b(k_0, k)
{\hat{\vec k}}\cdot\boldsymbol{\gamma}
\ee
with $\hat{\vec  k}=\vec k/|\vec k|$ (again
neglecting the possibility of color superconductivity).

Defining $\Sigma_\pm(k_0,k)\equiv b(k_0, k) \pm \,a(k_0, k)$,
a natural decomposition of the fermion self-energy and propagator
\index{fermion propagator} \index{fermion self-energy}
$S^{-1}= -{\not\! k} + \Sigma$ is given by
\bea\label{SIGL}
\gamma_0\Sigma(k_0, {\bf  k})&=&\Sigma_+(k_0,k)
\Lambda_+(\hat {\vec k})\,-\,\Sigma_-(k_0,k)
\Lambda_-(\hat {\vec k}), \\
\gamma_0 S^{-1} (k_0, {\bf  k})&=&\Delta_+^{-1}(k_0,k)
\Lambda_+(\hat {\vec k}) \,+\,\Delta_-^{-1}(k_0,k)\Lambda_-(\hat {\vec k})
\eea
with $\Delta_\pm^{-1} \equiv - [k_0\mp(k+\Sigma_\pm)]$
and spin matrices
\bea
\Lambda_{\pm}(\hat {\vec k})&\equiv& \frac{1 \pm \gamma^0
 \bgamma\cdot\hat{\vec k}}{2},\qquad \Lambda_++\Lambda_-=1,\\
\Lambda_{\pm}^2&=&\Lambda_{\pm},
\qquad \Lambda_+\Lambda_-\,=\,\Lambda_-\Lambda_+=0,
\qquad {\rm Tr} \Lambda_\pm = 2, 
\eea
projecting onto spinors whose chirality is equal ($\Lambda_+$),
or opposite ($\Lambda_-$), to their  helicity.

In the HTL approximation where $|k_0|,|{\bf  k}| \ll {\rm max}(T,\mu)$,
the fermion self-energy has been first calculated by Klimov 
\cite{Klimov:1981ka}
as
\be\label{SigmaHTL}
\Sigma^{\rm HTL}_\pm(k_0,k)\,=\,{\hat M^2\over k}\,\left(1\,-\,
\frac{k_0\mp k}{2k}\,\log\,\frac{k_0 + k}{k_0 - k}
\right)\ee
where $\hat M^2$ is the plasma frequency for fermions,
i.e., the frequency of long-wavelength ($k\to 0$) fermionic
excitations \index{plasma frequency!fermionic}
\be\label{MF}
\hat M^2 
={g^2 C_f\over 8}\left(T^2+{\mu^2\over \pi^2}\right).\ee
($C_f=(N^2-1)/2N$ in SU($N$) gauge theory, and $g^2C_f\to e^2$ in QED.)

\begin{figure}
%\centerline{ 
\includegraphics[viewport =                           %%%%
0 120 540 570,scale=0.38]{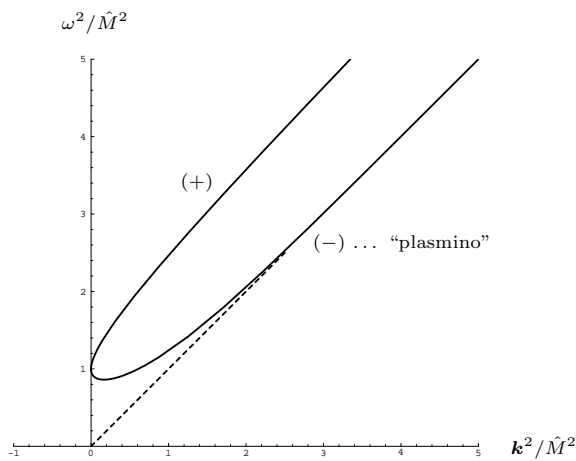} %} 

\bp\scriptsize
%\put(0,120){\fbox{\footnotesize quadratic scales!}}
\put(190,14){$\vec k^2/\hat M^2$}
\put(20,175){$\omega^2/\hat M^2$}
\put(65,115){$(+)$}
\put(115,92){$(-)$ \ldots { ``plasmino''}}
%\qbezier(71,46)(71,25)(100,45)
%\put(71,46){\vector(0,1){3}}
%\put(102,47){\footnotesize ``plasmino dip''}
\ep
\caption{The location of the zeros of $\Delta_\pm^{-1}$ in the
HTL approximation in quadratic scales.
The additional collective modes of branch $(-)$ (``plasminos'')
has a minimum of $\omega$ at $\omega/\hat M\approx 0.93$ and
$|\vec k|/\hat M\approx 0.41$ and approaches the light-cone for
large momenta, but with exponentially vanishing residue.
The regular branch approaches a mass hyperboloid (here a
straight line parallel to the diagonal) with asymptotic mass
$\sqrt2 \hat M$ \index{asymptotic thermal mass}
\label{Figf}}
\end{figure}

This leads to two separate branches of dispersion laws of
fermionic quasi-particles $\Delta_\pm(\omega,k)^{-1}=0$
carrying particle and hole quantum numbers respectively \cite{Weldon:1989ys,Pisarski:1989wb}. As shown in Fig.~\ref{Figf},
the $(-)$-branch, which is occasionally nicknamed ``plasmino'',
exhibits a curious dip reminiscent of the dispersion law of
rotons in liquid helium.\index{plasmino} \index{plasmino dip}

As we have seen in Sect.~\ref{giHTL}, gauge dependences start at
order $T$ in the high-temperature expansion. While the HTL result
(\ref{SigmaHTL}) is completely gauge independent, gauge parameter
dependences enter at subleading order. The
gauge dependence identity for the fermion propagator
\index{gauge dependence identities}
is, however, somewhat simpler than that for the gluon propagator.
The singularities in the fermion propagator can be summarized by
\be\label{detS}
 \det\!\( S^{-1}_{\sigma\bar \varrho} \)=0\ee
where $\sigma,\bar \varrho$ are spinor indices.
The gauge dependence identity thus takes the form
\be\label{ddetS}
\delta\det\( S^{-1}_{\sigma\bar \varrho} \)\!(k)
=-\det\( S^{-1}_{\sigma\bar \varrho} \)\!(k)\,
[\delta X^\tau_{,\tau}(k)+\delta X^{\bar \tau}_{,\bar \tau}(k)]\ee
where the two expressions within the square brackets
are Dirac traces of the diagrams appearing within the
braces in Fig.~\ref{FigdSigma}.

The conclusion of gauge independence of the solutions of (\ref{detS})
can now be reached by essentially the same arguments as those
for the gauge boson propagator (with the exception that now
there are no additional,
gauge-dependent kinematical poles like those arising in the
projection onto mode $B$ in (\ref{dDB})). The only obstruction
to gauge independence comes from singularities of 
$\delta X^\tau_{,\tau}(k)$ and $\delta X^{\bar \tau}_{,\bar \tau}(k)$.
If mass-shell singularities from massless gauge modes are
avoided by infrared regularization or finite volume, only
the singularities of the ghost propagator need to be considered.
Again, the latter are generically different from those leading to
the now fermionic quasi-particles since ghosts do not have
HTL self-energies $\sim T^2$.

We have thus seen that all the singularities of the fermion
propagator as well as those of the $A$- and $B$-branch of
the gluon propagator (with some exceptions for the latter)
are gauge-fixing independent.
On the other hand, residues (if those singularities are simple poles
at all) are not protected and may be gauge dependent; even the
nature of the singularity may be different from gauge to gauge,
as is well known to be
the case already for the electron propagator in zero-temperature QED
\cite{BogS:QF}.

%%%%% res %%%%%%%%%%%%%%%%

\section{Quasiparticles %at NLO 
in HTL Perturbation Theory}
\index{HTL quasi-particles}

We have already seen that loop order in bare perturbation theory
is not a good expansion parameter for calculating corrections
to quasi-particle properties at soft scales $\sim gT$.
Technically what happens is that the HTL contributions to
one-loop vertex functions are of the same order of magnitude
as their tree-level counterparts for external momenta $k\sim gT$:
\be
\quad { \Gamma_{,N}^{\rm HTL}} \sim g^N { T^2} { k^{2-N}}
\sim g^{N-2} { k^{4-N}} 
\sim {\partial ^N\CL\over \partial A^N}\Big|_{ k\sim gT}.
\ee
Therefore all HTL contributions need to be resummed in
Feynman diagrams that are sensitive to the soft regime $k\sim gT$.

Since HTL's are the leading contributions from hard momenta
$k \sim T$, this can be understood as the transition from
the bare Lagrangian to an effective,
Wilson-renormalized one for $ k\sim gT$,
$\CL\to \CL+\CL^{\rm HTL}$. $\CL^{\rm HTL}$ is the
effective Lagrangian containing all the HTL diagrams
and arises from integrating out all hard modes.

Soon after the identification of all the HTL's of QCD in 
\cite{Frenkel:1990br,Braaten:1990mz}, 
it has been found that, formally, $\CL^{\rm HTL}$ has a
comparatively simple and manifestly gauge-invariant
integral representation 
\cite{Taylor:1990ia,Braaten:1992gm,Frenkel:1992ts}%,Brandt:1993mj}
\index{HTL effective action}
\bea\CL^{\rm HTL} &=& \hat M^2 \int {d\O_v\over 4\pi} \bar \psi \g^\mu
{{ v_\m}\over { v}\cdot D(A)} \psi \nonumber\\
&& -{3\over 2}\omega^2_{pl.} {\rm tr} \int {d\O_v\over 4\pi} F^{\m\a}
{{ v_\a} { v^\b} \over  ({ v}\cdot D_{adj.}(A))^2} F_{\m\b}
\label{HTLeffL}\eea
with $\hat M^2$ the fermionic plasma frequency given in (\ref{MF}) and 
$\omega^2_{pl.}=\frac13\Pi^{\rm HTL}_\mu{}^\mu$ the more
familiar one of the gauge bosons
(cf. (\ref{PiHTLmumu})).
In this integral representation
$ v=(1,\vec v)$ is a light-like 4-vector, i.e.\ with $ \vec v^2=1$, 
and its
spatial components are averaged over by $ d\O_v$. 
$ v$ is the remnant of the hard plasma constituents'
momenta $p^\mu \sim T v^\mu$, namely their light-like
4-velocity, and the overall scale $T$ has %been
combined with the coupling constant to form the scale of
thermal masses, $\hat M, \omega_{pl.}\sim gT$.

The HTL effective Lagrangian (\ref{HTLeffL}) is
manifestly gauge invariant and moreover
gauge independent ($\hat M$ and $\omega_{pl.}$ do not
depend on the gauge fixing parameters used to integrate out
the hard modes). It is non-local
and 
Hermitian only in a Euclidean form, i.e.\
prior to analytic continuation to
real time/frequencies. It has cuts which
physically correspond to the phenomenon of \ind{Landau damping}.
The equations of motions associated with (\ref{HTLeffL})
can also be obtained from kinetic theory, which is extremely
useful to gain further physical insight 
\cite{Blaizot:1993zk,Blaizot:1994be,Kelly:1994dh,Blaizot:2001nr}.
There is also a noteworthy connection to Chern-Simons
theory \cite{Efraty:1992gk,Efraty:1993pd,Nair:1993rx}.

Using (\ref{HTLeffL}) as an effective theory for soft scales $\sim gT$
means that the bare propagators are to be replace by those of
HTL quasi-particles, and these have
infinitely many nonlocal vertices.
E.g., the three-gluon vertex becomes
\bea
{\Gamma^{abc}_{\mu\nu\varrho}}^{\rm cl+HTL}(k,q,r)
&=&\I g f^{abc}\Bigl\{g_\mn(k-q)_\varrho
 + {\rm cycl.}\nonumber\\&+&3\o^2_{\rm pl.}\int {d\O_v\over 4\pi}
v_\mu v_\nu v_\varrho \left[ {r_0\over  k\cdot v
\, r\cdot v}-{q_0\over  k\cdot v \, q\cdot v} \right] \Bigr\}.
\eea
In QCD, there are \ind{HTL vertices} for any number of external
gluons and up to two quark lines, whereas in QED, where
$v\cdot D_{\rm adj.}(A)) \to v\cdot \partial $ in (\ref{HTLeffL}),
there is  ``only'' an HTL photon self-energy $\Pi_\mn$, an HTL
fermion self-energy $\Sigma$,
and vertices involving two fermions and an arbitrary number of photons.

While the effective Lagrangian (\ref{HTLeffL}) is gauge invariant
and gauge independent in its entirety, NLO corrections won't be so.
However, as we have seen in the previous section, the positions
of singularities of
the effective (quasi-particle) propagators are protected against
gauge dependences by the identities (\ref{dDA}), (\ref{dDB}), and
(\ref{ddetS}).

\subsection{Long-Wavelength Plasmon Damping}

\index{plasmon!damping} \index{plasmon}
The first such correction to be calculated by means of
the HTL-resummed perturbation theory was the damping rate
of long-wavelength plasmons\footnote{For kinematical reasons, 
there should be no difference between 
spatially transverse and longitudinal gluonic quasi-particles
(cf. Fig.~\ref{Figg}), since with $\vec k\to0$ one can no longer
tell the one from the other. However, the limit $\vec k\to0$
involves infrared problems (see further below), and there
are even explicit
calculations \cite{Abada:1997vm,Abada:1998ue}
that claim to find obstructions to this equality,
which are however refuted by the recent work of \cite{Dirks:1999uc}.
}
\index{HTL quasi-particles!NLO corrections}
from the shift of the pole of the gluon propagator
at $\vec k=0$ from
$\omega=\omega_{\rm pl.}^{\rm HTL}$ $\to$ 
$\omega=\omega_{\rm pl.}-\I \gamma(\vec k=0)$
with the result \cite{Braaten:1990it}
\be\label{gamma6}
\gamma(\vec k=0)=+6.635\ldots{g^2NT\over 24\pi}=
0.264 \sqrt{N}{ g}\,\omega_{\rm pl.}^{\rm HTL},
\ee
implying the existence of 
weakly damped plasmons for $g\ll 1$. 

In QCD, where
one is %rather 
interested in the range $g\sim 1$,
one finds that the existence of plasmons as quasi-particles
requires that $g$ is significantly less than 2.2,
so real QCD is on the borderline of having identifiable
long-wavelength quasi-particles.\index{plasmon}

The corresponding quantity for fermionic quasi-particles has been
calcula\-ted in \cite{Kobes:1992ys,Braaten:1992gd}
with a comparable result: weakly damped long-wavelength
fermionic quasi-particles in 2- or 3-flavor QCD require that
$g$ is significantly less than 2.7.

\subsection{NLO Correction to Gluonic Plasma Frequency}
\index{plasma frequency}

In \cite{Schulz:1994gf}, Schulz has calculated
also the real part of the NLO
contribution to the gluon polarization tensor 
in the limit of $\vec k\to0$ which determines the
NLO correction to the gluonic plasma frequency.

The original power-counting arguments of \cite{Braaten:1990mz}
suggested that besides one-loop diagrams with HTL-resummed
propagators and vertices, there could be also contributions
from two-loop diagrams to relative order $g$. 
The explicit (and lengthy) calculation
of \cite{Schulz:1994gf} showed that those contribute only
at order $g^2 \ln(1/g)$ rather than $g$, and
the NLO plasma frequency
in a pure-glue plasma was obtained as
\be\label{NLOmpl}
\omega_{\rm pl.}=\omega_{\rm pl.}^{\rm HTL} \left[ 1-
0.09 \sqrt{N}{ g} \right].
\ee

In this particular result, HTL-resummed perturbation theory turns
out to give a moderate correction to the leading-order HTL value
even for $g\sim 1$; see however below.

While the calculations leading to (\ref{gamma6}) and (\ref{NLOmpl})
contain some interesting physics, in the following we shall go into
more detail only for a couple of more tractable cases, which
nonetheless will turn out to involve a number of salient
points.

\subsection{NLO Correction to the Non-Abelian Debye Mass}
\index{Debye mass}

The poles of the (gluon) propagator do not only give
the dispersion law of quasi-particles, but also the
screening of fields with frequencies below the plasma
frequency and in particular of static fields.
Below the plasma frequency, there are poles for $\vec k^2 < 0$,
as displayed in Fig.~\ref{Figg}, corresponding in configuration space to
exponential fall-off with (frequency-dependent)
screening mass $\sqrt{|\vec k^2|}$, i.e.\ screening
length $1/\sqrt{|\vec k^2|}$.

In the static case, branch $A$ of the gluon propagator describes
the screening of (chromo-)magnetostatic fields. 
While there is a finite screening length as long as $\omega>0$,
the $A$-branch of the HTL propagator becomes unscreened in
the static limit.
Whereas in QED, a ``\ind{magnetic mass}'' is forbidden by gauge invariance
\cite{Fradkin:1965,Blaizot:1995kg}, some sort of 
entirely non-perturbative magnetic mass
is expected in non-Abelian gauge theories in view of severe
infrared problems caused by the self-interactions of
magnetostatic gluons \cite{Polyakov:1978vu,Linde:1980ts,Gross:1981br}.

Branch $B$, on the other hand, contains the information about
screening of (chromo-)electric fields as generated by
static charges (Debye screening). The \ind{Debye mass} given by
the leading-order HTL propagator is $\hat m_D=\sqrt{3}\omega_{pl.}$.
The determination of its NLO correction has a history that
is at least as long as the plasmon (damping) puzzle, for it
starts already with (ultra-relativistic) QED.

Customarily, the \ind{Debye mass} (squared) has been {\em defined} as
the infrared limit $\Pi_{00}(\omega=0,k\to0)$, which indeed
is correct at the HTL level, cf.\ (\ref{PiBHTL}) and (\ref{PiHTL00}).

In QED, this definition has the advantage of being directly
related to a derivative of the thermodynamic pressure, so that
the higher-order terms known from the latter determine those
of $\Pi_{00}^{\rm QED}(\omega=0,k\to0)$ through \cite{Fradkin:1965,Kap:FTFT}
\be\label{PiQED00}
\Pi%^{\rm QED}
_{00}(0,k\to0)\Big|_{\mu=0}
=e^2{\partial ^2P\over \partial \mu^2}\Big|_{\mu=0}={e^2T^2\over 3}
\left(1-{3e^2\over 8\pi^2}+{\sqrt3 e^3\over 4\pi^3}
+\ldots\right)
\ee
This result is gauge independent because in QED all of $\Pi_\mn$ is.

In the case of QCD, there is no such relation. In fact,
one expects $\delta m_D^2/\hat m_D^2$ $\sim g$ rather than $g^3$
because of gluonic self-interactions and Bose enhancement.
The calculation of this quantity should be much easier than
the dynamic ones considered above, because in the static
limit the HTL effective action collapses 
to just the local, bilinear HTL \ind{Debye mass} term,
\be\label{HTLeffLstatic}
\CL^{\rm HTL} \stackrel{\rm static}{\longrightarrow}
-{1\over2} \hat m_D^2 {\rm tr} A_0^2.
\ee
This is also
gauge invariant, because $A_0$ behaves like an adjoint scalar under
time-independent gauge transformations.
Resummed perturbation theory thus boils down to a resummation
of the HTL Debye mass in the electrostatic propagator, which
is what had been done already since long \cite{Gell-Mann:1957,Kap:FTFT}.

Using this simple (``ring'') resummation\index{ring resummation}
in QCD, one finds however the gauge dependent result \cite{Toimela:1985ht}
\be\label{Pi0000}
\Pi_{00}(0,0)/m_D^2 = 1+{ \alpha}{N\over 4\pi}\sqrt{6\over 2N+N_f}g 
\ee
where $\alpha$ is the gauge parameter of general covariant gauge (which
coincides with general \ind{Coulomb gauge} in the static limit).
\index{covariant gauges}

This result was interpreted as meaning that the non-Abelian Debye
mass could not be obtained in resummed perturbation theory 
\cite{Nadkarni:1986cz} or that one should use a physical gauge
instead \cite{Kajantie:1985xx,Kap:FTFT}. In particular, temporal axial
\index{temporal gauge}
gauge was put forward, because in this gauge there is, like in
QED, a linear relationship between electric field strength
correlators and the gauge propagator. However, because
static ring resummation clashes with \ind{temporal gauge},
inconclusive and contradicting results were obtained by
different authors \cite{Kajantie:1982hu,Furusawa:1983gb,Kajantie:1985xx},
and in fact one cannot do without vertex resummations if
one wants to be consistent there \cite{Baier:1994et,Peigne:1995dn}.
But be that as it may be, switching to the chromoelectric field strength
correlator is not good enough, for it is gauge variant
and its infrared limit is equally gauge dependent \cite{Rebhan:1994mx}.

On the other hand, in view of the 
\ind{gauge dependence identities} discussed in
the previous section, the gauge dependence of (\ref{Pi0000}) is no longer
surprising. Gauge independence can only be expected ``on-shell'',
which here means $\omega=0$ but $\vec k^2\to-\hat m_D^2$.
%At HTL level, there is no difference to $|\vec k|=0$, since 
%$\Pi^{\rm HTL}_{00}(0,k)$ is constant.

Indeed, the exponential fall-off of the electrostatic propagator
is determined by the position of the singularity of $\Delta_B(0,k)$,
and not simply by its infrared limit.
This implies in particular that one should use a different definition
of the \ind{Debye mass} already in QED, despite the gauge independence
of (\ref{PiQED00}), namely \cite{Rebhan:1993az}
\be\label{mDpoledef}
m^2_D=\Pi_{00}(0,k)\Big|_{\vec k^2\to-m_D^2}.
\ee

For QED (with massless electrons), the \ind{Debye mass} is thus
not given by (\ref{PiQED00}) but rather as
\bea m^2_D&=&\Pi_{00}(0,k\to0)+\bigl[
%\underbrace{
\Pi_{00}(0,k)\big|_{k^2=-%\hat 
m^2_D}-\Pi_{00}(0,k\to0)
%}_{\makebox[0pt]{$
%{e^2 \vec k^2\over 6\pi^2}\left[\ln{\tilde\mu\over \pi T}
%+\g_E-{4\over 3}\right]_{\vec k^2=-\hat m^2_D}
%+O(e^4)$}}
\bigr]\nonumber\\
&=&{e^2T^2\over 3}
\biggl(1-{3e^2\over 8\pi^2}+{\sqrt3 e^3\over 4\pi^3}
+\ldots -{e^2 \over 6\pi^2}[\ln{\tilde\mu\over \pi T}+\g_E-{4\over 3}]
+\ldots \biggr)
\label{mDQED}
\eea
where $\tilde\mu$ is the renormalization scale of
the momentum subtraction scheme,\footnote{The slightly different
numbers in the terms $\propto e^4T^2$ 
quoted in \cite{Blaizot:1995kg,LeB:TFT}
pertain to the minimal subtraction (MS) scheme.}
i.e.\
$\Pi_\mn(k^2=-\tilde\mu^2)|_{T=0}=0$. Since 
$de/d\ln\tilde\mu=e^3/(12\pi^2)+O(e^5)$,
(\ref{mDQED}) is a renormalization-group invariant result for
the Debye mass in hot QED, which (\ref{PiQED00})
obviously failed to be.

In QCD, where gauge independence is not automatic,
the dependence on the gauge fixing parameter $\alpha$
is another indication that (\ref{Pi0000}) is
the wrong definition. For (\ref{mDpoledef})
we need the full momentum dependence of
the correction $\delta\Pi_{00}(k_0=0,{\vec k})$ to
$\Pi^{\rm HTL}_{00}$.
Since only the electrostatic
mode needs to be dressed, this is not difficult
to obtain \cite{Rebhan:1993az}:
\bea
\delta\Pi_{00}(k_0=0,{\vec k})
&=&\underbrace{\textstyle g\hat m_DN\sqrt{\frac6{2N+N_f}}}_{g^2T}\int
\frac{d^{3-2\varepsilon}p}{(2\pi)^{3-2\varepsilon}}\nonumber\\
&\times& \biggl\{\frac1{{\vec p}^2+\hat m_D^2}+\frac1{{\vec p}^2}
+\frac{4\hat m_D^2-({\vec k}^2+\hat m_D^2)
[3+2{\vec p}{\vec k}/{\vec p}^2]}{{\vec p}^2
%\underbrace
{[(\vec p+\vec k)^2+\hat m_D^2]}%_{\ddot\frown}
}
\nonumber\\
&&+{ \alpha} 
%\underbrace
{({\vec k}^2+\hat m_D^2)}%_{\ddot\smile}
\frac{{\vec p}^2+2{\vec p}{\vec k}}{{\vec p}^4{%\underbrace
{[(\vec p+\vec k)^2+\hat m_D^2]}%_{\ddot\frown}
}}
\biggr\}.
\label{dPi00QCD}
\eea

In accordance with the \ind{gauge dependence identities}, the last term
shows that gauge independence holds algebraically
for $\vec k^2=-\hat m_D^2$. %, on ``screening mass shell''.
On the other hand, on this ``screening mass shell'', where
the denominator term 
$[(\vec p+\vec k)^2+\hat m_D^2] \to [\vec p^2+2\vec p\vec k]$, we
encounter IR-singularities. In the $\alpha$-dependent term,
they are such that they produce a divergent factor $1/[\vec k^2+m_D^2]$
so that the gauge dependences no longer disappear even on-shell.
This is, however, the very same problem that had to be solved
in the above case of the plasmon damping in \ind{covariant gauges}. Introducing
a temporary infrared cut-off (e.g., finite volume), does not
modify the factor $[\vec k^2+m_D^2]$ in the numerator but
defuses the dangerous denominator. Gauge independence thus
holds for all values of this cut-off, which can be sent to zero in the end.
The gauge dependences are thereby identified as belonging to
the (infrared divergent) residue.

The third term in the curly brackets, however, remains
logarithmically singular on-shell as the infrared cut-off is to be
removed. In contrast to the $\alpha$-dependent term, closer
inspection reveals that these singularities are coming from
the massless magnetostatic modes and not from unphysical massless
gauge modes. 

At HTL level, there is no (chromo-)magnetostatic screening, but, as we
have mentioned, one expects some sort of such screening to be generated
non-perturbatively in the static sector of hot QCD at
the scale $g^2T \sim gm_D$
\cite{Polyakov:1978vu,Linde:1980ts,Gross:1981br}.
\index{magnetic mass}

While this singularity prevents evaluating (\ref{dPi00QCD}) in full,
the fact that this singularity is only logarithmic allows one
to extract the leading term of (\ref{dPi00QCD}) under the
assumption of an effective cut-off at $p \sim g^2T$ as \cite{Rebhan:1993az}
\be\label{mDlng}
{\delta m^2_D\over \hat m^2_D}=
\frac{N}{2\pi}\sqrt{\frac6{2N+N_f}}\,g\,\ln\frac1g+O(g).
\ee

The $O(g)$-contribution, however, is sensitive to the physics
of the magnetostatic sector at scale $g^2T$, and is completely
non-perturbative in that all loop order $\ge 2$ are
expected to contribute with equal importance.

Because of the undetermined $O(g)$-term in (\ref{mDlng}), one-loop
resummed perturbation theory only says that for sufficiently small $g$,
where $O(g\ln(1/g))\gg O(g)$,
there is a {\em positive} correction to the Debye
mass of lowest-order perturbation theory following from
the pole definition (\ref{mDpoledef}), and that it is gauge
independent.

On the lattice, the static gluon propagator of pure SU(2)
gauge theory at high temperature \index{gluon propagator!lattice}
has been studied in various gauges \cite{Heller:1997nq,Cucchieri:2001tw}
with the result that the electrostatic propagator is exponentially
screened with a screening mass that indeed appears to be gauge independent
and which is about 60\% larger than the leading-order \ind{Debye mass}
for temperatures $T/T_c$ up to about $10^4$.

In \cite{Rebhan:1994mx}, an estimate of the $O(g)$
contribution to (\ref{mDlng}) has been made using the crude
approximation of a simple massive propagator for the magnetostatic
one, which leads to
\be\label{mDlnm}
{\delta m^2_D\over \hat m^2_D}=
\frac{N}{2\pi}\sqrt{\frac6{2N+N_f}}\,g\,\left[\ln{2m_D\over m_m}-\frac12
\right].
\ee
On the lattice one finds
strong gauge dependences of the magnetostatic
screening function, but the data
are consistent with an over-all exponential
behaviour corresponding to $m_m \approx 0.5 g^2T$ in all gauges
\cite{Heller:1997nq,Cucchieri:2000cy}. Using this number
in a self-consistent evaluation of (\ref{mDlnm}) gives
an estimate for $m_D$ which is about 20\% larger than
the leading-order value for $T/T_c=10\ldots 10^4$.

This shows that there are strong non-perturbative contributions
to the Debye screening mass $m_D$ even at very high temperatures.
Assuming that these are predominantly of order $g^2T$, one-loop resummed
perturbation theory (which is as far as one can get)
is able to account for about 1/3 of
this inherently non-perturbative physics already, if one introduces a 
simple, purely phenomenological magnetic screening mass.

\subsubsection{Other Non-Perturbative Definitions of the Debye Mass}
\index{Debye mass}

A different approach to studying Debye screening non-perturbatively
without the complication of gauge fixing is to consider spatial
correlation functions of appropriate gauge-invariant operators
such as those of the \ind{Polyakov loop}
\be
L(\vec x)={1\over N} {\rm Tr}\, {\cal P} \exp \left\{
-\I g \int_0^\beta d\tau \, A_0(\tau,\vec x)\right\}.
\ee
The correlation of two such operators is related to the free energy
of a quark-antiquark pair \cite{McLerran:1981pb}. In lowest
order perturbation theory this is given by the square of
a Yukawa potential with screening mass $\hat m_D$
\cite{Nadkarni:1986cz}; at
one-loop order one can in fact identify contributions
of the form (\ref{mDlnm}) if one assumes magnetic screening
\cite{Braaten:1994pk,Rebhan:1994mx}, but there is the problem that 
through higher loop orders the
large-distance behaviour becomes 
dominated by the magnetostatic modes and their 
lightest bound states \cite{Braaten:1995qx}.

In \cite{Arnold:1995bh}, Arnold and Yaffe have proposed to use Euclidean
time reflection symmetry to distinguish electric and magnetic
contributions to screening, and have given a prescription to
compute the sublogarithmic contribution of order $g^2T$ to $m_D$
nonperturbatively. This has been carried out in 3-d lattice simulations
for SU(2) \cite{Kajantie:1997pd,Laine:1997nq} as well as for
SU(3) \cite{Laine:1999hh}. The Debye mass thus defined shows
even larger deviations from the lowest-order perturbative results
than that from gauge-fixed lattice propagators. E.g.,
in SU(2) at $T=10^4 T_c$ this deviation turns out to be over 100\%,
while in SU(3) the dominance of $g^2T$
contributions is even more pronounced.

Clearly, (resummed) perturbation theory is of no use here for
any temperature of practical interest. However, the magnitude
of the contributions from the completely nonperturbative magnetostatic
sector depends strongly on the quantity considered. It is
significantly smaller in the definition of the \ind{Debye mass} through
the exponential decay of gauge-fixed
gluon propagators, which, as we have seen, 
leads to smaller screening masses on the
lattice (and gauge-independent ones, too, apparently). 
In quantities where the barrier in perturbation
theory arising from the magnetostatic sector occurs at higher
orders, HTL-resummed perturbation theory should be in much
better shape, and we shall find some support for this
%more optimistic attitude 
further below.

%%%%%%%%%%%%%%%%%%%%%%%%%%%%%%%%%%

\subsection{Dynamical Damping and Screening}
\index{HTL quasi-particles!damping}

A logarithmic sensitivity to the nonperturbative physics
of the magnetostatic sector has in fact been encountered
early on also in the calculation of damping of a heavy fermion 
\cite{Pisarski:1989vd}, and more generally of hard
particles \cite{Lebedev:1990ev,Lebedev:1991un,Burgess:1992wc,Rebhan:1992ca}.
It also turns out to occur for
soft quasi-particles as soon as they are propagating
\cite{Pisarski:1993rf,Flechsig:1995sk} and not just stationary
plasma oscillations.

\begin{figure}
\includegraphics[width=7truecm]{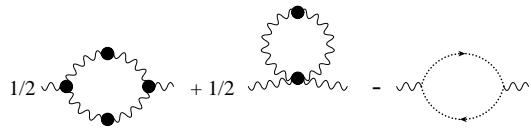}
\caption{One-loop diagrams in HTL-resummed perturbation theory.
HTL-resummed quantities are marked with a blob
 \label{Figdpi}}
\end{figure}

Because this logarithmic sensitivity arises only if one internal
line of (resummed) one-loop diagrams is static, the
coefficient of the resulting $g\ln(1/g)$-term is almost
as easy to obtain as in the case of the Debye mass, even though
the external line is non-static and soft, requiring
HTL-resummed vertices (see Fig.~\ref{Figdpi}).

The infrared singularity arises (again) from the dressed one-loop
diagram with two propagators, one of which is magnetostatic and
thus massless in the HTL approximation, and the other of the same
type as the external one, so only the first diagram in Fig.~\ref{Figdpi}
is relevant. The dressed 3-vertices in it are needed only
in the limit of one leg being magnetostatic and having zero momentum.
Because of the gauge invariance of HTL's, these are determined
by the HTL self-energies through a differential Ward identity, e.g.\
\index{Ward identities}
\be
\hat\Gamma_{\mu\nu\varrho}(k;-k;0)=
-{\partial \over \partial k^\varrho}\hat\Pi_{\mu\nu}(k)
\ee
for the 3-gluon vertex (color indices omitted).

Comparatively simple algebra gives \cite{Flechsig:1995sk}
\be
\delta\Pi_I(k) \simeq -g^2 N 4 \vec k^2 [1+\partial _{\vec k^2}\Pi_I(k)]^2 
{ \mathscr S_I(k)},\qquad {I=A,B}
\label{dPiI}
\ee
where
\be{ \mathscr S_I(k)} := T\int{d^3p\over (2\pi)^3}{ 1\over \vec p^2}
{-1\over (k-p)^2-\Pi_I(k-p)} \Big|_{k^2=\Pi_I(k), p^0=0}
\ee
and the logarithmic (mass-shell) singularity arises because
$(k-p)^2-\Pi_I(k-p) \to -\vec p^2+2\vec p\vec k-\Pi_I(k-p)
+\Pi_I(k)\, { \sim |\vec p|} $ as $k^2 \to \Pi_I(k)$.

The IR-singular part of ${\mathscr S}_I(k)$ is given by
\bea
{\mathscr S}_I(k)&=&T\int{d^3p\over (2\pi)^3}{1\over \vec p^2}
{1\over \vec p^2-2\vec p\vec k+\Pi_I(k-p)-\Pi_I(k)-\I \varepsilon}\nonumber\\
&\simeq& T [1+\partial _{\vec k^2}\Pi_I(k)]^{-1}\int{d^3p\over (2\pi)^3}
{1\over \vec p^2}{1\over \vec p^2-2\vec p\vec k-\I \varepsilon}\nonumber\\
&=&T[1+\partial _{\vec k^2}\Pi_I(k)]^{-1} \int_{ \lambda}^\infty
{dp\over p}{1\over 2|\vec k|}\ln{p+2|\vec k|-\I \varepsilon \over
  p-2|\vec k|-\I \varepsilon} \label{calSI}
\eea
where in the last line we have inserted an 
IR cutoff $\lambda\ll gT$ for the $p$-integral
in order to isolate the singular behaviour.

One finds that (\ref{calSI}) has a singular imaginary part
for 
propagating modes,
\be{\mathscr S}_I(k)\simeq \I{T\over 8\pi|\vec k|}
[1+\partial _{\vec k^2}\Pi_I(k)]^{-1} 
\ln{|\vec k|\over \lambda} + O(\lambda^0) 
\ee
for $\vec k^2>0$ ($\vec k$ real), and a singular real part
in screening situations where
$|\vec k|^2 \to - \kappa^2$, $\kappa \in \mathbb R$ 
(i.e., $\vec k$ imaginary):
\be{\mathscr S}_I(k)\simeq
+{T\over 8\pi\kappa}[1+\partial _{\vec k^2}\Pi_I(k)]^{-1} 
\ln{\kappa\over \lambda} + O(\lambda^0) .
\ee

So from one and the same expression we can see that logarithmic
IR singularities arise whenever $|\vec k|\not=0$, leading to
IR singular contributions to damping or (dynamical) screening,
depending on whether $\omega>0$ or $<0$ and thus
$\vec k^2>0$ or $<0$.
The case $\vec k=0$ is IR-safe, because (\ref{dPiI}) is proportional
to $\vec k^2$, while
\be\label{SIk0}
{\mathscr S}_I(k)\longrightarrow {T\over 4\pi^2 \lambda}
[1+\partial _{\vec k^2}\Pi_I(k)]^{-1} + O\({T|\vec k|\over \lambda^2}\)
\quad \mbox{for $\vec k \to 0$.}
\ee

This shows that there is a common origin for the infrared sensitivity
of screening and damping of HTL quasi-particles.
The perturbatively calculable coefficients of the resulting
$g \ln(1/g)$-terms are in fact beautifully simple:
For the damping of moving quasi-particles one obtains
\cite{Pisarski:1993rf,Flechsig:1995sk}
\index{HTL quasi-particles!damping}
\be\label{Damping}%
\quad\gamma_I(|\vec k|) 
\simeq {g^2NT\over 4\pi}%\underbrace
{|\vec k|[1+\partial _{\vec k^2}\Pi_I(k)]\over 
\omega(|\vec k|)[1-\partial _{\omega^2}\Pi_I(k)]}%_{v_I} 
\ln{1\over g} 
\equiv{g^2NT\over 4\pi}v_I(|\vec k|) \ln{1\over g} 
\ee
where $v_I(|\vec k|)$ is the group
velocity of mode $I$ (which vanishes at $\vec k=0$). 
The IR-sensitive NLO correction to
screening takes its simplest form when formulated as \cite{Flechsig:1995sk}
\be
\quad\delta\kappa^2_I(\omega)={g^2NT\over 2\pi}\kappa_I(\omega)
\( \ln{1\over g}+ O(1) \)
\ee
where $\kappa_I(\omega)$ is the inverse screening length of mode $I$
at frequency $\omega<\omega_{pl.}$ (which in the static limit
approaches the \ind{Debye mass} and
perturbatively vanishing magnetic mass, resp., while
approaching zero for both modes as $\omega\to\omega_{pl.}$).

A completely analogous calculation for the fermionic modes (for
which there are no screening masses) gives
\be\label{gpm}
\gamma_\pm(|\vec k|)={g^2C_FT\over 4\pi}|v_\pm|(|\vec k|) 
\( \ln{1\over g} + O(1) \).
\ee
for $|\vec k|>0$. 
The group velocity $v_\pm$ equals $\pm\frac13$ in the limit
$(|\vec k|)\to 0$, and increases monotonically towards $+1$ for
larger momenta (with a zero for the $(-)$-branch at $|\vec k|/\hat M
\approx 0.41$). 
For strictly $|\vec k|=0$, the IR sensitivity in fact disappears
because (\ref{gpm}) is no longer valid for $|\vec k|\ll\lambda$,
but one has $\gamma_\pm(|\vec k|)|_{\rm sing.}
\propto g^2 T |\vec k|/\lambda$ instead.
Thus $\gamma_\pm(0)$ is calculable at order $g^2T$ in HTL-resummed
perturbation theory, and has been calculated in 
\cite{Kobes:1992ys,Braaten:1992gd}.

The fermionic result (\ref{gpm}) applies in fact equally to QED, for
which one just needs to replace $g^2C_F\to e^2$.
This is particularly disturbing as QED does not allow
a non-zero \ind{magnetic mass} as IR cutoff, and
it has been conjectured that the damping $\gamma\sim g^2T$ or $e^2T$
itself might act as an effective IR cutoff 
\cite{Lebedev:1990ev,Lebedev:1991un,Pisarski:1993rf,Altherr:1993ti},
which however led to further difficulties \cite{Peigne:1993ky}.
The solution for QED was finally
found by Blaizot and Iancu \cite{Blaizot:1996hd,Blaizot:1997az,Blaizot:1997kw}
who showed that there the fermionic modes undergo over-exponential damping
in the form $\E^{-\gamma t} \to \E^{-{e^2\over 4\pi}%|v_\pm|
T t \ln( %|v_\pm|
\omega_{\rm pl.}t)}$ (for $v\to1$),
so finite time %acts as the IR cut-off here. 
is the actual IR cut-off.
The fermion propagator
has in fact no simple quasi-particle pole, but nevertheless a sharply
peaked spectral density.
%In fact, for $t \sim \gamma^{-1}$, the typical life-time of
%excitations, the perturbative result (\ref{gpm}) turns out
%to be correct after all.

In non-Abelian
theories, on the other hand, one does expect static (chro\-mo-)\-magnetic
field to have finite range, and lattice results do confirm
this, so the above estimates may be
appropriate after all, at least for sufficiently weak coupling.

%%%%%%%%%%%%%%%

\subsection{NLO Corrections to Real Parts of Dispersion Laws}
%excitations with 
%$\9k^2/\omega_{\rm pl.}^2 \gg 1 $}

The above analysis has identified the imaginary parts of the
dispersion laws to be sensitive to non-perturbative IR physics
except at $\vec k=0$ and where the group velocity vanishes (which
includes one further point at $|\vec k|\not=0$ for the fermionic plasmino
branch). %Formulated in positive terms, 
On the other hand, the real parts
of the dispersion laws of gluonic and fermionic quasi-particles
are IR-safe in NLO HTL-resummed perturbation theory.
However, such calculations are tremendously difficult, and
only some partial results exist so far 
in QCD \cite{Flechsig:1995uz,Flechsig:1998mn}.

In the following, we shall restrict our attention to
the case $\vec k^2/\omega_{\rm pl.}^2 \gg 1 $ and consider
the two branches of the gluon/photon propagator in turn.
In both cases, interesting physics will be seen to be
contained in the NLO corrections.

\subsubsection{Longitudinal Plasmons}\index{plasmon}

For momenta $\vec k^2 \gg \omega_{\rm pl.}^2$, the longitudinal
plasmon branch approaches the light-cone, as can be seen in
Fig.~\ref{Figg}. From 
$k^2=\Pi_B^{\rm HTL}(k)$ and (\ref{PiBHTL}) one finds
\be
\omega^2_B(|\vec k|) \to 
\vec k^2\(1+4\vec k^2 {\rm e}^{-6\vec k^2/(e^2 T^2)}\)
\ee
with $e^2 = g^2(N+N_f/2)$ in QCD, so the light-cone is
approached exponentially. If one also calculates the residue,
one finds that this goes to zero at the same time, and exponentially so, too. 

Instead of QCD, we shall consider the
analytically tractable case of massless \ind{scalar electrodynamics}
as
%Concerning NLO calculations in HTL-resummed perturbation theory,
%massless \ind{scalar electrodynamics} is 
a simple toy model
with at least some similarities to the vastly more complicated QCD case
in that in both theories there are
bosonic self-interactions.
There are however no HTL vertices in scalar electrodynamics, 
which makes it possible
to do complete momentum-dependent NLO calculations \cite{Kraemmer:1995az}.

Comparing HTL values of and NLO corrections to $\Pi_B$, one finds
that as $k^2\to0$
there are collinear singularities in both:
\index{collinear singularities}
\be\Pi_B^{\rm HTL}(k)/k^2 \to {3\over 2}\omega_{\rm pl.}^2 
{ \ln{\vec k^2\over k^2}}
\ee
diverges logarithmically\footnote{This is in fact the technical reason
why the longitudinal branch approaches the light-cone exponentially
when $\vec k^2 \gg \omega_{\rm pl.}^2$.}, whereas
\be\label{dPiBNLOlc}
\delta\Pi_B/k^2 \to -e\mu_{\rm sc.th.}^2{ |\vec k|\over {\sqrt{k^2}}}
\ee
(with $\mu_{\rm sc.th.}\propto e T$ the thermal mass of the scalar).
Because (\ref{dPiBNLOlc})
diverges stronger than logarithmically, one has
$ \delta\Pi_B > \Pi_B^{\rm HTL}$ eventually as $ k^2\to 0$.
Clearly, this leads to a breakdown of perturbation theory
in the immediate neighbourhood of the light-cone (${k^2/|\vec k|^2}
\lesssim (e/\ln{1\over e})^2$), which
this time is not caused by the massless magnetostatic modes, but
rather by the massless hard modes contained in the HTL's.

\begin{figure}
\includegraphics[%bb=72 280 540 520,
viewport = 0 0 240 440,clip,width=5.5truecm]{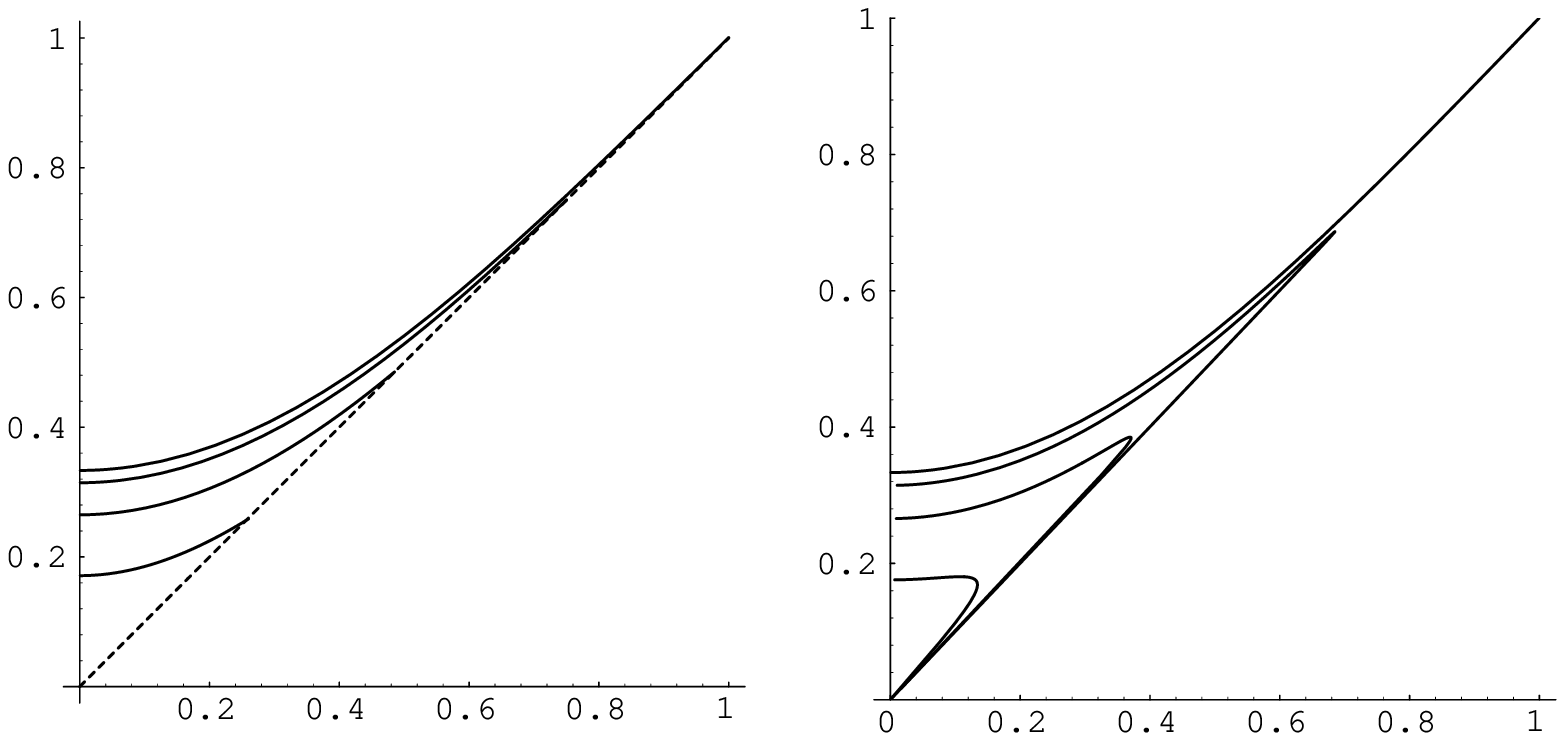}

\vspace{-140pt}
\bp
\scriptsize
\put(10,157){$\omega/(eT)$}
\put(150,20){$|\vec k|/(eT)$}
%\put(20,100){$e=0,0.3,1,2$}
\ep
\caption{ The longitudinal \ind{plasmon} branch of 
\ind{scalar electrodynamics}
including NLO corrections to the HTL result. The upper of the
four lines gives the HTL result and the lines below correspond
to NLO corrections with $e=0.3$, 1, and 2, respectively.
The latter three lines cross the light-cone such that the
phase velocity starts to exceed 1, but with group velocity $<1$
throughout. In the space-like region, the plasmon modes are
damped by \ind{Landau damping}, which is strong except in the immediate
neighborhood of the light-cone, where it is suppressed by
a factor of $\exp\{-e\sqrt{\vec k/[8(\vec k-\omega)]}\}$
\label{Figpllc}}
\end{figure}

However, a self-consistent gap equation for the scalar thermal
mass implies that also the hard scalar modes have a thermal
mass $\sim eT$. Including this by extending the resummation
of the scalar thermal mass to hard internal lines
renders $\Pi_B$ regular up to and including the light-cone one obtains
\be\label{PiBk20}
\lim\limits_{k^2\to0}{\Pi_B^{\rm resum.}\over  k^2}
= {e^2T^2\over  3\vec k^2}
  \Bigl[ \underbrace{{ \ln{ 2T\over
  \mu_{\rm sc.th.}}}}_{\ln{4\over e}}+{1\over  2}
-\gamma_E+{\z'(2)\over  \z(2)} \Bigr]
%  - {e^2T\mu_{\rm sc.th.}\over  2\pi \vec k^2}
+\ldots
\ee
The finiteness of (\ref{PiBk20}) makes it possible that there is
now a solution to the dispersion law with $k^2=0$
at 
$\vec k^2/(e^2T^2)={1\over 3}\ln{2.094\ldots \over e}%-{e\over 4\pi}
+O(e)$. Because all \ind{collinear singularities} have disappeared,
continuity implies that there are also solutions for
space-like momenta $k^2<0$, so the longitudinal plasmon branch pierces
the light-cone, having group velocity $v<1$ throughout, though,
as shown in Fig.~\ref{Figpllc}.
While at HTL level, the strong \ind{Landau damping} at $k^2<0$
switches on discontinuously, it now does so smoothly
through an extra factor $\exp[-e\sqrt{|\vec k|/[8(|\vec k|-\omega)]}]$,
removing the longitudinal plasmons through over-damping
for $(|\vec k|-\omega)/|\vec k| \gtrsim e^2$.

So the \ind{collinear singularities} of HTL-resummed perturbation
theory on the light-cone were associated with a slight but
nevertheless qualitative change of the spectrum of longitudinal
plasmons: instead of being time-like throughout and 
existing for higher momenta, albeit with
exponentially small and decreasing
residue and effective mass, they become space-like at
a particular point $|\vec k| \sim eT\ln{1\over e}$ and expire
through \ind{Landau damping} soon thereafter.

This phenomenon is in fact known to occur in non-ultrarelativistic 
($T<m_e$) QED \cite{Tsytovich:1961}, and has been considered in the case of
QCD in a little-known paper by Silin and Ursov \cite{Silin:1988js},
who speculated that it may lead to Cherenkov phenomena in the
quark-gluon plasma.

In QCD, the situation is in fact much more complicated. Under the
assumption that the \ind{collinear singularities} are removed solely by the
resummation of \index{asymptotic thermal mass}
asymptotic gluonic and fermionic thermal masses
in hard internal lines, the value of $|\vec k|$
where longitudinal plasmons turn space-like
has been calculated in \cite{Kraemmer:1995az}. \index{plasmon}
For a pure-glue plasma, it reads
\be\label{k2critQCD}
\vec k^2_{\rm crit.}=g^2T^2[\ln{1.48\ldots\over g}+O(g)].
\ee
Such an extended resummation can in fact be related to
an improved and still gauge-invariant version of the
HTL effective action \cite{Flechsig:1996ju}, however
it may well be that damping effects are of equal importance
here (in contrast to scalar electrodynamics), so that
(\ref{k2critQCD}) may not be complete. A similar
unsolved problem occurs in the calculation of the
production rate of real, non-thermalized photons in
a quark-gluon plasma from HTL-resummed perturbation
theory \cite{Baier:1994zb,Aurenche:1996is,Aurenche:1996sh}.

Taken at face value, (\ref{k2critQCD}) 
would imply that propagating longitudinal plasmons
do no longer exist for $g\gtrsim 1.48$, and a negative $O(g)$
contribution would even lower this bound. \index{plasmon}

%%%%%%%%%%%%%%%%%%%%%%%

\subsubsection{Energetic Quarks and Transverse Gluons and Their Role
in Self-Consi\-stent Thermodynamics}

At high momenta
$ \vec k^2/\omega_{\rm pl.}^2 \gg 1 $ the additional collective
modes of longitudinal plasmons and ``plasminos'' disappear.
\index{plasmon} \index{plasmino}
At HTL level, they do so because the residues of the corresponding
poles in the gluon and quark propagators vanish exponentially,
whereas at NLO, as we have just seen, they cross the light-cone
and die from strong \ind{Landau damping}. The remaining transverse
gluonic and normal quark modes on the other hand approach
asymptotic mass hyperboloids.
For transverse gauge bosons the asymptotic thermal photon/gluon mass
of the HTL approximation reads
\be\label{mAHTLas}
\Pi_A^{\rm HTL} \to { m_\infty^2} = {e^2 T^2\over 6}
\ee
($e^2=(N+N_f/2)g^2$ for gluons);
\index{asymptotic thermal mass}
whereas in the case of fermions we have
\be
2|\vec k|\Sigma_+^{\rm HTL} \to { 2\hat M^2}
\ee
with $\hat M$ the HTL fermionic plasma frequency given by (\ref{MF}).

These results remain the correct LO ones even for $\omega,|\vec k| \sim T$,
because the light-cone values of $\Pi_A$ and $\Sigma_+$ are
identical to their HTL/HDL values there and do not depend on
the HTL approximation that $\omega,|\vec k| \ll T$ 
\cite{Kraemmer:1990dr,Flechsig:1996ju}.

The asymptotic thermal masses play an interesting role in
self-consistent (approximations to) thermodynamics 
\cite{Blaizot:1999ip,Blaizot:1999ap,Blaizot:2000fc}:
The LO ($\propto g^2$) interaction piece of the \index{entropy}
{\em entropy density} can be expressed in terms of the
light-cone values of the various self-energies and thus
the asymptotic thermal masses. \index{asymptotic thermal mass}
E.g., in the pure-glue case, 
the $g^2$-contribution to the entropy density reads
\bea
s^{(2)} &=& -(N^2-1)\int\!\!{d^3k\,d\omega\over (2\pi)^3}
{\partial n(\o)\over \partial T} 
{\rm sgn}(\o) \d(\o^2-k^2)
{\rm Re}\Pi_T(\o,k) \nonumber\\
&=& -{(N^2-1)\over 6}{ m_\infty^2} T=-{N(N^2-1)\over 36}g^2T^3.
\eea
Fermionic contributions give similarly
\be
s^{(2)}_f=-{NN_f\over 6}{ M_\infty^2} T,
\ee
possibly with nonzero chemical potential $\mu$.
With nonzero $\mu$, one can also consider the
\ind{quark density}, which likewise is determined
by the asymptotic mass: \index{asymptotic thermal mass}
\be 
n^{(2)}_f=-{NN_f\over 2\pi^2}{ M_\infty^2}\mu.
\ee
Up to a $T$- and $\mu$-independent integration constant,
\ind{entropy} and quark densities determine the complete thermodynamical
potential, and the above formula give nice, universal formulae
for the LO interaction terms.

Remarkably, also the NLO interaction term $\propto g^3$ can be directly
related to the properties of HTL/HDL quasiparticles. The
so-called plasmon term of the thermodynamic potential $\propto g^3$
is usually understood as arising from the
resummation of the static \ind{Debye mass}, which
needs to be kept only in the zero modes of the electrostatic
gluon propagator. The resulting coefficient of the order-$g^3$
contribution to the thermodynamic potential turns out, however,
to have an uncomfortably large value,\footnote{The same holds true
for the order-$g^5$ contribution which has been calculated for
QCD in \cite{Arnold:1994ps,Arnold:1995eb,Zhai:1995ac,Braaten:1996jr}.}
and appears to spoil completely
the convergence of perturbation theory for all temperatures
smaller than some $10^5 T_c$.

While it is correct that all that is needed for a calculation
of the thermodynamic potential through order $g^3$ is to approximate
quarks and gluons by their vacuum spectral densities except
for the one massive electrostatic mode \cite{Arnold:1993rz}, 
this is clearly a cruder
approximation than that of HTL-resummed propagators which
contain a lot of physics beyond Debye screening.

In \cite{Blaizot:1999ip,Blaizot:1999ap,Blaizot:2000fc} it
has been shown recently that in a self-consistent formulation
of the thermodynamic potentials \ind{entropy} and density one can
find a {\em real-time} description of those using quasi-particles
which at soft momenta are described by the HTL effective
propagators and at hard momenta by their light-cone limits
and NLO corrections thereof. Doing so, it turns out that
a larger part (up to $3\over 4$) of the (soft) plasmon effect $\propto g^3$
comes from the NLO corrections to the {\em hard} asymptotic masses,
\index{asymptotic thermal mass}
reflecting a massive\footnote{Pun intended.} 
reorganization of usual (Debye-screened)
perturbation theory:
\bea\label{s3glue}
{3\over 4}s^{(3)}&=&{3\over 4}(N^2-1){\hat m_D^3\over 3\pi}\nonumber\\&=&
-(N^2-1)\int\!\!{d^3k\over (2\pi)^3}\,{1\over k}\,
{\partial n(k)\over \partial T}\,
{
\underbrace{{\rm Re}\,\delta \Pi_T(\o=k)}_{\textstyle 
\delta m_\infty^2(k)}}\;\;
\eea
(in the case of pure glue).

$\delta m_\infty^2$ in HTL-resummed perturbation theory is
a non-local (momentum-depedent)
correction, which is infrared safe and thus calculable.
Through the relation (\ref{s3glue}) one can 
define the average correction
\be\label{dminfty}
\bar\delta m_\infty^2=-{1\over 2\pi}g^2NT\hat m_D
\ee
which has a remarkably simple form.
Similarly, for fermions one finds
\be
\bar\delta M_\infty^2=-{1\over 2\pi}g^2C_fT\hat m_D.
\ee

Now, numerically, this correction is uncomfortably large:
\be{\bar\delta m_\infty^2\over m_\infty^2}=1-{\sqrt{3N}\over \pi}g\approx 1-g
\ee 
(pure glue) so
that perturation theory seems to become completely useless for
$g \gtrsim 1$, i.e., $\alpha_s \gtrsim 0.1$.

However, a very
similar problem arises already in simple scalar $\phi^4$ theory.
If one considers the large-$N$ limit of the iso-vector 
${\rm O}(N)$ $g^2\vec \phi^4$ theory, one can write down
an exact gap equation of the form \cite{Dolan:1974qd,Drummond:1997cw}
\be\label{scgap}
m^2=12g^2 \int\!{d^3k\over (2\pi)^3}{n(\sqrt{k^2+m^2})\over \sqrt{k^2+m^2}}
+{3m^2\over 4\pi^2}\(\ln{m^2\over \bar\mu^2}-1\)
\ee
whose solution
has a perturbative expansion beginning as
\be\label{m2scnlo}
m^2=g^2T^2(1-{3\over \pi}g+\ldots),
\ee
which happens to have the same $O(g)$ coefficent as the QCD
result (\ref{dminfty}), and which likewise
gives nonsense such as tachyonic thermal masses for $g\gtrsim 1$.

\begin{figure}
\includegraphics[bb = 70 180 540 540,width=7.5truecm]{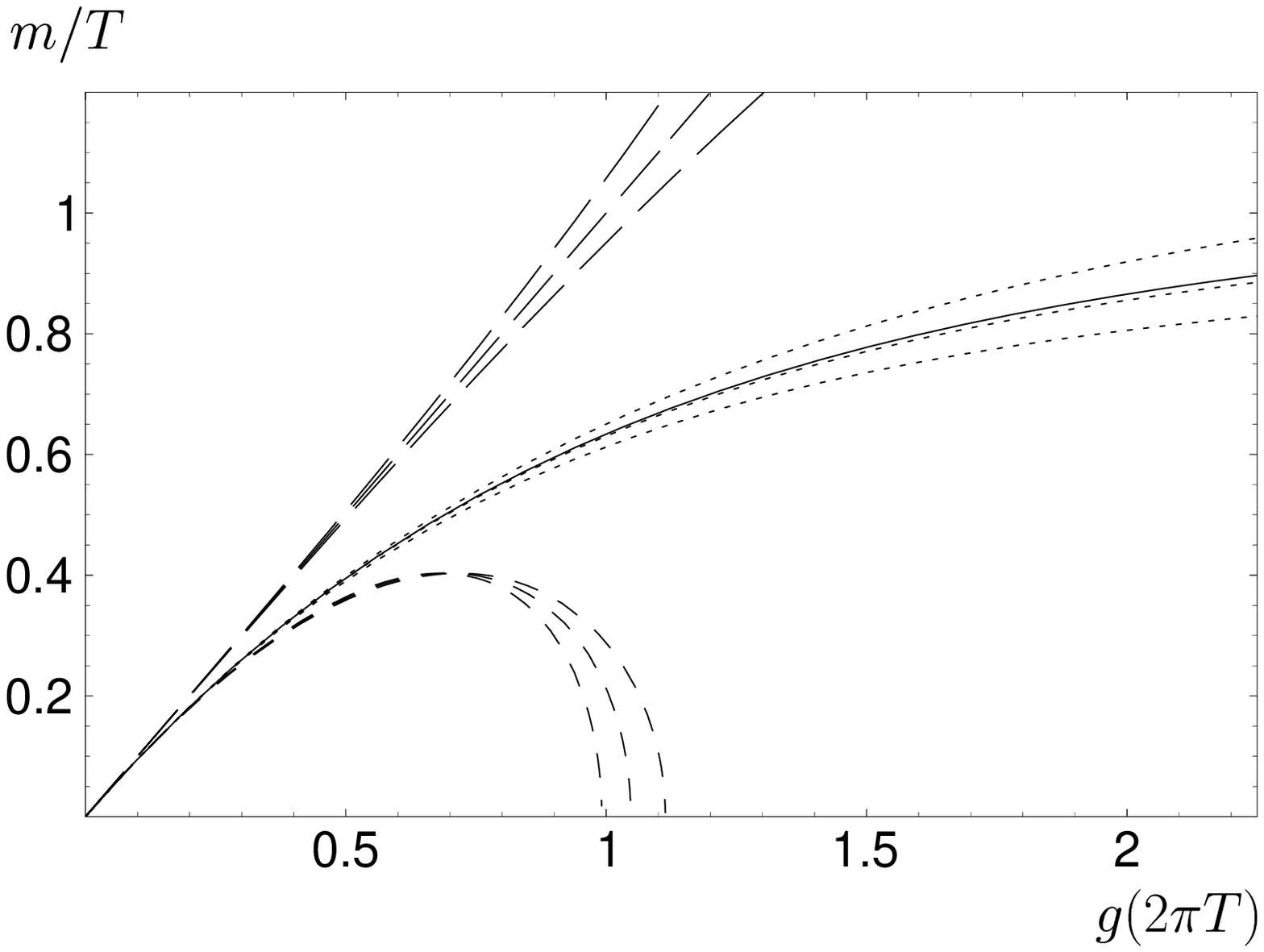}
\bp\footnotesize
\put(-4,117){\scriptsize $\leftarrow$ exact}
\put(-35,124){ASC}
\put(-145,110){LO}
\put(-135,40){NLO}
\put(-85,37){\fbox{$\bar\mu=\pi T,2\pi T,4\pi T$}}
\ep
\caption{Various approximations to the thermal mass of a scalar boson
in large-$N$ $\ph^4$ theory: %as a function
%of the coupling defined at $\bar\mu=2\pi T$:
leading-order HTL (LO), next-to-leading order (NLO) as given
by \ref{m2scnlo}), and the approximately self-consistent (ASC)
gap equation (\ref{ASCgap}), which is perturbatively equivalent to NLO.
The $\overline{\mbox{MS}}$
renormalization scale
is varied by a factor of 2 about $\bar\mu=2\pi T$
\label{Figdm}}
\end{figure}

However, if one instead writes down an
approximate gap equation by expanding in powers of $m/T$ and dropping
terms of order $(m/T)^2\sim g^2$:
\be\label{ASCgap}
m^2=g^2T^2-{3\over \pi}g^2Tm,
\ee
then one finds that the solution to this simple quadratic equation in
$m$ gives a function $m(g)$ that is perturbatively equivalent to
(\ref{m2scnlo}), but does not go mad for $g\gtrsim 1$. On the contrary,
for the standard choice of renormalization scale $\bar\mu=2\pi T$
in $\overline{\mbox{MS}}$, it gives a remarkably accurate
approximation of the solution to the full gap equation (\ref{scgap}),
as is shown in Fig.~\ref{Figdm}.

\begin{figure}
\includegraphics[bb = 70 180 540 540,width=7.5truecm]{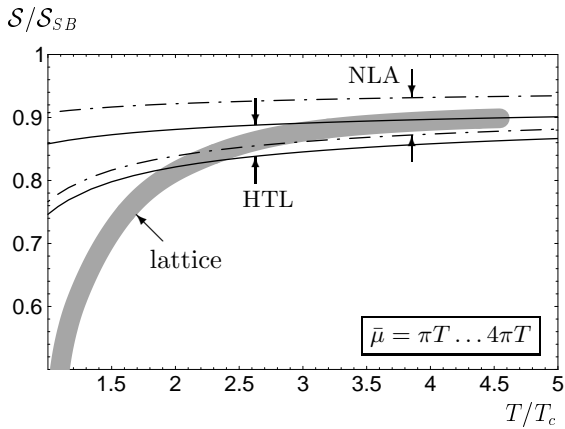}
\bp\footnotesize
\put(-80,37){\fbox{$\bar\mu=\pi T \ldots 4\pi T$}}
\put(-155,75){\vector(-1,1){10}}
\put(-160,65){\normalsize lattice}
%\put(-180,-15){\normalsize Boyd et al 1996, Okamoto et al 1999}
\put(-85,135){NLA}
\put(-61,140){\vector(0,-1){10}}
\put(-61,105){\vector(0,1){10}}
\put(-125,88){HTL}
\put(-120,129){\vector(0,-1){10}}
\put(-120,97){\vector(0,1){10}}
\ep
\caption{Comparison of results from approximately self-consistent
thermodynamics based on the HTL propagator and a next-to-leading
approximation (NLA) using the analogue of (\ref{ASCgap}) for
the asymptotic gluon mass correction (\ref{dminfty}) with lattice
data from \cite{Boyd:1996bx}. The gray band gives roughly
the lattice data with their errors. The analytical results
are given with two boundaries corresponding to a variation of
the $\overline{\mbox{MS}}$
renormalization scale $\bar\mu$ from $\pi T$ to $4\pi T$
 \label{FigSg}}
\end{figure}

Implementing analogous ``approximately self-consistent''
gap equations for the hard modes, a non-perturbative, UV finite and
gauge-invariant approximation
to \ind{entropy} and density of hot QCD has been proposed in
\cite{Blaizot:1999ip,Blaizot:1999ap,Blaizot:2000fc}.
It is perturbatively equivalent to conventional Debye-resummed
perturbation theory but goes beyond the latter in incorporating
all of the collective phenomena contained in HTL propagators
as well as NLO effects in their asymptotic masses.
\index{asymptotic thermal mass}
When compared to available lattice data 
\cite{Boyd:1996bx} (see Fig.~\ref{FigSg} for
the pure-glue case), remarkable agreement is found
down to temperatures $\sim 3 T_c$.
By contrast, conventionally resummed perturbation theory 
at order $g^3$ leads
to ${\cal S}/{\cal S}_{\rm SB} > 1$ for all but exceedingly
high temperatures.

An optimistic conclusion one could draw from this is that the
transition to gluonic and quark quasi-particles is able to
absorb a large part of the strong elementary interaction into
the spectral properties of the former, and that, at least in
infrared-safe situations, these quasi-particles have
comparatively weak residual interactions even in QCD at temperatures
a few times above the transition temperature.

Let me recall that even in the infrared-unsafe case of NLO corrections
to the Debye mass, the self-consistent NLO result (\ref{mDlnm}) using a
phenomenological magnetic mass gives the qualitatively correct
result of a substantially increased electric mass, while underestimating
the magnitude of the increase
by a factor of 3 when compared to lattice simulations
of chromoelectrostatic propagators.

\section{Conclusions}

Let us summarize the findings that are of specific interest to
a perturbative formulation of non-Abelian gauge theories at
finite temperature and/or density: 

The leading-order results for self-energies and vertices
in a high-temper\-ature/density expansion, the so-called
hard thermal (dense) loops, form a gauge-invariant
and gauge-independent effective action, which is the basis
for a systematic perturbative expansion in powers of $g$
(rather than $g^2$), as long as one does not
run into the perturbative barrier formed by the
completely non-perturbative self-interacting chromomagnetostatic
modes.

Beyond leading order, gauge dependences appear in all
Green functions of the fundamental fields. In particular
the propagators which are expected to carry information on gluonic or
fermionic quasi-particles depend on gauge-fixing parameters.
The gauge dependence identities that we have discussed above
imply, however, that under certain conditions the location of
the singularities which define the dispersion laws of these
quasi-particles are gauge-independent, though not e.g.\
residues or even the type of the singularities, which need not
be simple poles.

Already at NLO, screening lengths and damping constants
are logarithmically infrared sensitive to the nonperturbative
magnetostatic sector, with the exception of zero 3-momentum.
Infrared-safe quantities are also the real corrections to the
dispersion laws, which in the case of longitudinal plasmons
(and also of the plasmino branch of fermions) lead to
a finite 3-momentum range, and, at its upper end, to
space-like phase velocities. In the case of transverse gluonic
quasiparticles and the normal branch of fermionic ones,
the NLO corrections play an important role in self-consistent
formulations of thermodynamics (the equation of state).

Particularly in QCD, one faces the problem that corrections to
LO results are rather large for almost all values of the
coupling of interest. However, we have seen indications
that this poor convergence of thermal perturbation theory
may be overcome in approximately self-consistent 
reformulations.\footnote{There are alternative methods to
reorganize thermal perturbation theory which aim to improve its convergence.
%The perhaps simplest is the use of Pad\'e approximants
%\cite{Kastening:1997rg,Hatsuda:1997wf,Drummond:1997cw}.
A particularly interesting one is ``screened'' or
``optimized'' perturbation theory
\cite{Karsch:1997gj,Chiku:1998kd,Andersen:2000yj} which employs a single mass
parameter in a variational ansatz. In
\cite{Andersen:1999fw,Andersen:1999sf,Andersen:1999va}
an extension of this method to gauge theories has been proposed
which uses the HTL effective action uniformly at soft as well
as hard momenta with the thermal-mass prefactors $\omega_{\rm pl.}^2$ 
and $\hat M^2$ turned into
variational parameters. In contrast to the entropy-based approach,
this requires explicit dressed 2-loop contributions (involving
HTL vertices) in order to contain the correct %perturbative 
LO interaction coefficients in the thermodynamic pressure.}
Where those can be implemented, the picture of weakly interacting
quasi-particles even in strong interactions seems to have
some support from comparison with lattice data (where the latter
are available), and may remain valid down to a few times the
deconfinement phase transition temperature.
This picture, being primarily
set up in real Minkowski space, is complementary to 
lattice or dimensional reduction formulations, 
and allows (analytical) calculations from first principles
also where lattice gauge theory calculations are not (yet) feasible.
Its potentials, in particular when combined with results from
other nonperturbative approaches, are, in my opinion, not yet fully explored.

%\newpage
\section*{Acknowledgements}

I would like to thank all the organizers of the 40th Schladming Winter School
on ``Dense Matter'' for their efforts and
for having once again set up a marvellous meeting.
Furthermore, I would like to thank %in particular 
Jean-Paul Blaizot, Ian Drummond, Frithjof Flechsig, Ron Horgan,
Edmond Iancu, Randy Kobes, Ulli Kraemmer, Gabor Kunstatter,
Peter Landshoff, and Hermann Schulz for
past and present collaborations that provided a large part of the material
I had the privilege to present at this meeting.

%\bibliography{tft,tftpr,qft,grav,ar,books}

\begin{thebibliography}{100}

\bibitem{Abada:1998ue}
A.~Abada, O.~Azi: Phys. Lett. \textbf{B463}, 117 (1999)

\bibitem{Abada:1997vm}
A.~Abada, O.~Azi, K.~Benchallal: Phys. Lett. \textbf{B425}, 158 (1998)

\bibitem{Abbott:1981hw}
L.~F. Abbott: Nucl. Phys. \textbf{B185}, 189 (1981)

\bibitem{Altherr:1993ti}
T.~Altherr, E.~Petitgirard, T.~del Rio~Gaztelurrutia: Phys. Rev. \textbf{D47},
  703 (1993)

\bibitem{Andersen:1999fw}
J.~O. Andersen, E.~Braaten, M.~Strickland: Phys. Rev. Lett. \textbf{83}, 2139
  (1999)

\bibitem{Andersen:1999sf}
J.~O. Andersen, E.~Braaten, M.~Strickland: Phys. Rev. \textbf{D61}, 014017
  (2000)

\bibitem{Andersen:1999va}
J.~O. Andersen, E.~Braaten, M.~Strickland: Phys. Rev. \textbf{D61}, 074016
  (2000)

\bibitem{Andersen:2000yj}
J.~O. Andersen, E.~Braaten, M.~Strickland: Phys. Rev. \textbf{D63}, 105008
  (2001)

\bibitem{Arnold:1993rz}
P.~Arnold, O.~Espinosa: Phys. Rev. \textbf{D47}, 3546 (1993)

\bibitem{Arnold:1995bh}
P.~Arnold, L.~G. Yaffe: Phys. Rev. \textbf{D52}, 7208 (1995)

\bibitem{Arnold:1994ps}
P.~Arnold, C.-X. Zhai: Phys. Rev. \textbf{D50}, 7603 (1994)

\bibitem{Arnold:1995eb}
P.~Arnold, C.-X. Zhai: Phys. Rev. \textbf{D51}, 1906 (1995)

\bibitem{Aurenche:1996is}
P.~Aurenche, F.~Gelis, R.~Kobes, E.~Petitgirard: Phys. Rev. \textbf{D54}, 5274
  (1996)

\bibitem{Aurenche:1996sh}
P.~Aurenche, F.~Gelis, R.~Kobes, E.~Petitgirard: Z. Phys. \textbf{C75}, 315
  (1997)

\bibitem{Baier:1994et}
R.~Baier, O.~K. Kalashnikov: Phys. Lett. \textbf{B328}, 450 (1994)

\bibitem{Baier:1992mg}
R.~Baier, G.~Kunstatter, D.~Schiff: Nucl. Phys. \textbf{B388}, 287 (1992)

\bibitem{Baier:1992dy}
R.~Baier, G.~Kunstatter, D.~Schiff: Phys. Rev. \textbf{D45}, 4381 (1992)

\bibitem{Baier:1994zb}
R.~Baier, S.~Peign{\'e}, D.~Schiff: Z. Phys. \textbf{C62}, 337 (1994)

\bibitem{Bakshi:1963}
P.~M. Bakshi, K.~T. Mahanthappa: J. Math. Phys. \textbf{4}, 1 (1963)

\bibitem{Bernard:1974bq}
C.~W. Bernard: Phys. Rev. \textbf{D9}, 3312 (1974)

\bibitem{Blaizot:1993zk}
J.~P. Blaizot, E.~Iancu: Phys. Rev. Lett. \textbf{70}, 3376 (1993)

\bibitem{Blaizot:1994be}
J.~P. Blaizot, E.~Iancu: Nucl. Phys. \textbf{B417}, 608 (1994)

\bibitem{Blaizot:1996hd}
J.-P. Blaizot, E.~Iancu: Phys. Rev. Lett. \textbf{76}, 3080 (1996)

\bibitem{Blaizot:1997az}
J.-P. Blaizot, E.~Iancu: Phys. Rev. \textbf{D55}, 973 (1997)

\bibitem{Blaizot:1997kw}
J.-P. Blaizot, E.~Iancu: Phys. Rev. \textbf{D56}, 7877 (1997)

\bibitem{Blaizot:2001nr}
J.-P. Blaizot, E.~Iancu: \textit{The quark-gluon plasma: Collective dynamics
  and hard thermal loops}, hep-ph/0101103 (2001)

\bibitem{Blaizot:1995kg}
J.-P. Blaizot, E.~Iancu, R.~R. Parwani: Phys. Rev. \textbf{D52}, 2543 (1995)

\bibitem{Blaizot:1999ip}
J.~P. Blaizot, E.~Iancu, A.~Rebhan: Phys. Rev. Lett. \textbf{83}, 2906 (1999)

\bibitem{Blaizot:1999ap}
J.~P. Blaizot, E.~Iancu, A.~Rebhan: Phys. Lett. \textbf{B470}, 181 (1999)

\bibitem{Blaizot:2000fc}
J.~P. Blaizot, E.~Iancu, A.~Rebhan: Phys. Rev. \textbf{D63}, 065003 (2001)

\bibitem{BogS:QF}
N.~N. Bogoliubov, D.~V. Shirkov: \textit{Introduction to the theory of
  quantized fields} (Interscience Publishers, New York, 1959)

\bibitem{Boyd:1996bx}
G.~Boyd, et~al.: Nucl. Phys. \textbf{B469}, 419 (1996)

\bibitem{Braaten:1994pk}
E.~Braaten, A.~Nieto: Phys. Rev. Lett. \textbf{73}, 2402 (1994)

\bibitem{Braaten:1995qx}
E.~Braaten, A.~Nieto: Phys. Rev. Lett. \textbf{74}, 3530 (1995)

\bibitem{Braaten:1996jr}
E.~Braaten, A.~Nieto: Phys. Rev. \textbf{D53}, 3421 (1996)

\bibitem{Braaten:1990it}
E.~Braaten, R.~D. Pisarski: Phys. Rev. \textbf{D42}, 2156 (1990)

\bibitem{Braaten:1990az}
E.~Braaten, R.~D. Pisarski: Nucl. Phys. \textbf{B339}, 310 (1990)

\bibitem{Braaten:1990kk}
E.~Braaten, R.~D. Pisarski: Phys. Rev. Lett. \textbf{64}, 1338 (1990)

\bibitem{Braaten:1990mz}
E.~Braaten, R.~D. Pisarski: Nucl. Phys. \textbf{B337}, 569 (1990)

\bibitem{Braaten:1992gd}
E.~Braaten, R.~D. Pisarski: Phys. Rev. \textbf{D46}, 1829 (1992)

\bibitem{Braaten:1992gm}
E.~Braaten, R.~D. Pisarski: Phys. Rev. \textbf{D45}, 1827 (1992)

\bibitem{Burgess:1992wc}
C.~P. Burgess, A.~L. Marini: Phys. Rev. \textbf{D45}, 17 (1992)

\bibitem{Chiku:1998kd}
S.~Chiku, T.~Hatsuda: Phys. Rev. \textbf{D58}, 076001 (1998)

\bibitem{Cornwall:1985eu}
J.~M. Cornwall, W.-S. Hou, J.~E. King: Phys. Lett. \textbf{B153}, 173 (1985)

\bibitem{Cucchieri:2000cy}
A.~Cucchieri, F.~Karsch, P.~Petreczky: Phys. Lett. \textbf{B497}, 80 (2001)

\bibitem{Cucchieri:2001tw}
A.~Cucchieri, F.~Karsch, P.~Petreczky: \textit{Propagators and dimensional
  reduction of hot SU(2) gauge theory}, hep-lat/0103009 (2001)

\bibitem{Dewitt:1967ub}
B.~S. De{W}itt: Phys. Rev. \textbf{162}, 1195 (1967)

\bibitem{Dirks:1999uc}
M.~Dirks, A.~Niegawa, K.~Okano: Phys. Lett. \textbf{B461}, 131 (1999)

\bibitem{Dolan:1974qd}
L.~Dolan, R.~Jackiw: Phys. Rev. \textbf{D9}, 3320 (1974)

\bibitem{Drummond:1997cw}
I.~T. Drummond, R.~R. Horgan, P.~V. Landshoff, A.~Rebhan: Nucl. Phys.
  \textbf{B524}, 579 (1998)

\bibitem{Drummond:1999si}
I.~T. Drummond, R.~R. Horgan, P.~V. Landshoff, A.~Rebhan: Phys. Lett.
  \textbf{B460}, 197 (1999)

\bibitem{Efraty:1992gk}
R.~Efraty, V.~P. Nair: Phys. Rev. Lett. \textbf{68}, 2891 (1992)

\bibitem{Efraty:1993pd}
R.~Efraty, V.~P. Nair: Phys. Rev. \textbf{D47}, 5601 (1993)

\bibitem{Flechsig:1998mn}
F.~Flechsig: Nucl. Phys. \textbf{B547}, 239 (1999)

\bibitem{Flechsig:1996ju}
F.~Flechsig, A.~K. Rebhan: Nucl. Phys. \textbf{B464}, 279 (1996)

\bibitem{Flechsig:1995sk}
F.~Flechsig, A.~K. Rebhan, H.~Schulz: Phys. Rev. \textbf{D52}, 2994 (1995)

\bibitem{Flechsig:1995uz}
F.~Flechsig, H.~Schulz: Phys. Lett. \textbf{B349}, 504 (1995)

\bibitem{Fradkin:1965}
E.~S. Fradkin: Proc. Lebedev Inst. \textbf{29}, 1 (1965)

\bibitem{Frenkel:1990br}
J.~Frenkel, J.~C. Taylor: Nucl. Phys. \textbf{B334}, 199 (1990)

\bibitem{Frenkel:1992ts}
J.~Frenkel, J.~C. Taylor: Nucl. Phys. \textbf{B374}, 156 (1992)

\bibitem{Furusawa:1983gb}
T.~Furusawa, K.~Kikkawa: Phys. Lett. \textbf{B128}, 218 (1983)

\bibitem{Gelis:1999nx}
F.~Gelis: Phys. Lett. \textbf{B455}, 205 (1999)

\bibitem{Gell-Mann:1957}
M.~Gell-Mann, K.~A. Brueckner: Phys.Rev. \textbf{106}, 364 (1957)

\bibitem{Gross:1981br}
D.~J. Gross, R.~D. Pisarski, L.~G. Yaffe: Rev. Mod. Phys. \textbf{53}, 43
  (1981)

\bibitem{Hansson:1987un}
T.~H. Hansson, I.~Zahed: Nucl. Phys. \textbf{B292}, 725 (1987)

\bibitem{Hata:1980yr}
H.~Hata, T.~Kugo: Phys. Rev. \textbf{D21}, 3333 (1980)

\bibitem{Heinz:1987kz}
U.~Heinz, K.~Kajantie, T.~Toimela: Ann. Phys. \textbf{176}, 218 (1987)

\bibitem{Heller:1997nq}
U.~M. Heller, F.~Karsch, J.~Rank: Phys. Rev. \textbf{D57}, 1438 (1998)

\bibitem{Israel:1981}
W.~Israel: Physica \textbf{106A}, 204 (1981)

\bibitem{James:1990fd}
K.~A. James: Z. Phys. \textbf{C48}, 169 (1990)

\bibitem{James:1990it}
K.~A. James, P.~V. Landshoff: Phys. Lett. \textbf{B251}, 167 (1990)

\bibitem{Kajantie:1982hu}
K.~Kajantie, J.~Kapusta: Phys. Lett. \textbf{B110}, 299 (1982)

\bibitem{Kajantie:1985xx}
K.~Kajantie, J.~Kapusta: Ann. Phys. \textbf{160}, 477 (1985)

\bibitem{Kajantie:1997pd}
K.~Kajantie, et~al.: Phys. Rev. Lett. \textbf{79}, 3130 (1997)

\bibitem{Kalashnikov:1980cy}
O.~K. Kalashnikov, V.~V. Klimov: Sov. J. Nucl. Phys. \textbf{31}, 699 (1980)

\bibitem{Kap:FTFT}
J.~I. Kapusta: \textit{Finite-temperature field theory} (Cambridge University
  Press, Cambridge, UK, 1989)

\bibitem{Karsch:1997gj}
F.~Karsch, A.~Patk\'os, P.~Petreczky: Phys. Lett. \textbf{B401}, 69 (1997)

\bibitem{Keldysh:1964ud}
L.~V. Keldysh: Zh. Eksp. Teor. Fiz. \textbf{47}, 1515 (1964)

\bibitem{Kelly:1994dh}
P.~F. Kelly, Q.~Liu, C.~Lucchesi, C.~Manuel: Phys. Rev. \textbf{D50}, 4209
  (1994)

\bibitem{Klimov:1981ka}
V.~V. Klimov: Sov. J. Nucl. Phys. \textbf{33}, 934 (1981)

\bibitem{Kobes:1992ys}
R.~Kobes, G.~Kunstatter, K.~Mak: Phys. Rev. \textbf{D45}, 4632 (1992)

\bibitem{Kobes:1990xf}
R.~Kobes, G.~Kunstatter, A.~Rebhan: Phys. Rev. Lett. \textbf{64}, 2992 (1990)

\bibitem{Kobes:1991dc}
R.~Kobes, G.~Kunstatter, A.~Rebhan: Nucl. Phys. \textbf{B355}, 1 (1991)

\bibitem{Kraemmer:1990dr}
U.~Kraemmer, M.~Kreuzer, A.~Rebhan: Ann. Phys. \textbf{201}, 223 (1990)

\bibitem{Kraemmer:1995az}
U.~Kraemmer, A.~K. Rebhan, H.~Schulz: Ann. Phys. \textbf{238}, 286 (1995)

\bibitem{Kugo:1978zq}
T.~Kugo, I.~Ojima: Phys. Lett. \textbf{B73}, 459 (1978)

\bibitem{Kugo:1979eg}
T.~Kugo, I.~Ojima: Prog. Theor. Phys. \textbf{61}, 644 (1979)

\bibitem{Laine:1997nq}
M.~Laine, O.~Philipsen: Nucl. Phys. \textbf{B523}, 267 (1998)

\bibitem{Laine:1999hh}
M.~Laine, O.~Philipsen: Phys. Lett. \textbf{B459}, 259 (1999)

\bibitem{Landshoff:1992ne}
P.~V. Landshoff, A.~Rebhan: Nucl. Phys. \textbf{B383}, 607 (1992)

\bibitem{Landshoff:1993ag}
P.~V. Landshoff, A.~Rebhan: Nucl. Phys. \textbf{B410}, 23 (1993)

\bibitem{Landsman:1987uw}
N.~P. Landsman, C.~G. van Weert: Phys. Rept. \textbf{145}, 141 (1987)

\bibitem{LeB:TFT}
M.~{Le Bellac}: \textit{Thermal Field Theory} (Cambridge University Press,
  Cambridge, UK, 1996)

\bibitem{Lebedev:1990ev}
V.~V. Lebedev, A.~V. Smilga: Ann. Phys. \textbf{202}, 229 (1990)

\bibitem{Lebedev:1991un}
V.~V. Lebedev, A.~V. Smilga: Phys. Lett. \textbf{B253}, 231 (1991)

\bibitem{Linde:1980ts}
A.~D. Linde: Phys. Lett. \textbf{B96}, 289 (1980)

\bibitem{Matsubara:1955ws}
T.~Matsubara: Prog. Theor. Phys. \textbf{14}, 351 (1955)

\bibitem{Matsumoto:1983gk}
H.~Matsumoto, Y.~Nakano, H.~Umezawa: Phys. Rev. \textbf{D28}, 1931 (1983)

\bibitem{Matsumoto:1984au}
H.~Matsumoto, Y.~Nakano, H.~Umezawa: J. Math. Phys. \textbf{25}, 3076 (1984)

\bibitem{McLerran:1981pb}
L.~D. McLerran, B.~Svetitsky: Phys. Rev. \textbf{D24}, 450 (1981)

\bibitem{Nadkarni:1986cz}
S.~Nadkarni: Phys. Rev. \textbf{D33}, 3738 (1986)

\bibitem{Nadkarni:1988ti}
S.~Nadkarni: Phys. Rev. Lett. \textbf{61}, 396 (1988)

\bibitem{Nair:1993rx}
V.~P. Nair: Phys. Rev. \textbf{D48}, 3432 (1993)

\bibitem{Niegawa:1989dr}
A.~Niegawa: Phys. Rev. \textbf{D40}, 1199 (1989)

\bibitem{Peigne:1993ky}
S.~Peign{\'e}, E.~Pilon, D.~Schiff: Z. Phys. \textbf{C60}, 455 (1993)

\bibitem{Peigne:1995dn}
S.~Peign{\'e}, S.~M.~H. Wong: Phys. Lett. \textbf{B346}, 322 (1995)

\bibitem{Pisarski:1989wb}
R.~D. Pisarski: Nucl. Phys. \textbf{A498}, 423C (1989)

\bibitem{Pisarski:1989vd}
R.~D. Pisarski: Phys. Rev. Lett. \textbf{63}, 1129 (1989)

\bibitem{Pisarski:1993rf}
R.~D. Pisarski: Phys. Rev. \textbf{D47}, 5589 (1993)

\bibitem{Polyakov:1978vu}
A.~M. Polyakov: Phys. Lett. \textbf{B72}, 477 (1978)

\bibitem{Rajagopal:2000rs}
K.~Rajagopal, E.~Shuster: Phys. Rev. \textbf{D62}, 085007 (2000)

\bibitem{Rebhan:1987wp}
A.~Rebhan: Nucl. Phys. \textbf{B288}, 832 (1987)

\bibitem{Rebhan:1992ca}
A.~Rebhan: Phys. Rev. \textbf{D46}, 482 (1992)

\bibitem{Rebhan:1992ak}
A.~Rebhan: Phys. Rev. \textbf{D46}, 4779 (1992)

\bibitem{Rebhan:1993az}
A.~K. Rebhan: Phys. Rev. \textbf{D48}, 3967 (1993)

\bibitem{Rebhan:1994mx}
A.~K. Rebhan: Nucl. Phys. \textbf{B430}, 319 (1994)

\bibitem{Schulz:1994gf}
H.~Schulz: Nucl. Phys. \textbf{B413}, 353 (1994)

\bibitem{Schwinger:1961qe}
J.~Schwinger: J. Math. Phys. \textbf{2}, 407 (1961)

\bibitem{Silin:1988js}
V.~P. Silin, V.~N. Ursov: Sov. Phys. - Lebedev Inst. Rep. \textbf{5}, 43 (1988)

\bibitem{Taylor:1990ia}
J.~C. Taylor, S.~M.~H. Wong: Nucl. Phys. \textbf{B346}, 115 (1990)

\bibitem{Toimela:1985ht}
T.~Toimela: Z. Phys. \textbf{C27}, 289 (1985)

\bibitem{Tsytovich:1961}
V.~N. Tsytovich: Sov. Phys. (JETP) \textbf{13}, 1249 (1961)

\bibitem{Vilkovisky:1984st}
G.~A. Vilkovisky: Nucl. Phys. \textbf{B234}, 125 (1984)

\bibitem{Weldon:1982aq}
H.~A. Weldon: Phys. Rev. \textbf{D26}, 1394 (1982)

\bibitem{Weldon:1989ys}
H.~A. Weldon: Phys. Rev. \textbf{D40}, 2410 (1989)

\bibitem{Zhai:1995ac}
C.-X. Zhai, B.~Kastening: Phys. Rev. \textbf{D52}, 7232 (1995)

\end{thebibliography}
%\bibliographystyle{springer}

%INDEX%%%%%%%%%%%%%%%%%%%%%%%%%%%%%%%%%%%%%%%%%%%%%%%%%%%%%%%%%%%%%%%
% Please check with the editor of your book whether he plans to
% include a "mutual" subject index - if so, please code your entries
% in the standard syntax. For your own purposes you may print your
% "personal" index by using the following commands:
%
%\clearpage
%\addcontentsline{toc}{section}{Index}
%\flushbottom
%\printindex
%%%%%%%%%%%%%%%%%%%%%%%%%%%%%%%%%%%%%%%%%%%%%%%%%%%%%%%%%%%%%%%%%%%%%

\end{document}